\documentclass[prc,amsmath,twocolumn,superscriptaddress]{revtex4}

\usepackage{gensymb}
\usepackage{graphicx,color}
\usepackage{amssymb}
\usepackage{enumerate}
\usepackage{verbatim}
\usepackage{natbib}
\usepackage{amsmath}
\usepackage{float}
\usepackage{epsfig}
\usepackage{epstopdf}

\begin{document}
\newcommand {\nc} {\newcommand}
\nc {\Sec} [1] {Sec.~\ref{#1}}
\nc {\IR} [1] {\textcolor{red}{#1}} 
\nc {\IB} [1] {\textcolor{blue}{#1}}

\title{Constraining Transfer Cross Sections Using Bayes' Theorem}

\author{A.~E.~Lovell}
\email{lovell@nscl.msu.edu}
\affiliation{National Superconducting Cyclotron Laboratory, Michigan State University, East Lansing, MI 48824, USA}
\affiliation{Department of Physics and Astronomy, Michigan State University, East Lansing, MI 48824, USA}
\author{F.~M.~Nunes}
\email{nunes@nscl.msu.edu}
\affiliation{National Superconducting Cyclotron Laboratory, Michigan State University, East Lansing, MI 48824, USA}
\affiliation{Department of Physics and Astronomy, Michigan State University, East Lansing, MI 48824, USA}

\date{\today}


\begin{abstract}
\begin{description}
\item[Background:]   Being able to rigorously quantify the uncertainties in reaction models is crucial to moving this field forward. 
Even though Bayesian methods are becoming increasingly popular in nuclear theory, they have yet to be implemented and applied in reaction theory problems. 
\item[Purpose:]  The purpose of this work is to investigate, using Bayesian methods, the uncertainties in the optical potentials generated from fits to elastic-scattering data and the subsequent uncertainties in the transfer predictions.  We also study the differences in two reaction models where the parameters are constrained in a similar manner, as well as the impact of reducing the experimental error bars on the data used to constrain the parameters.
\item[Method:]  We use Bayes' Theorem combined with a Markov Chain Monte Carlo to determine posterior distributions for the parameters of the optical model, constrained by neutron-, proton-, and/or deuteron-target elastic scattering.  These potentials are then used to predict transfer cross sections within the adiabatic wave approximation or the distorted-wave Born approximation.  
\item[Results:]  We study a number of reactions involving deuteron projectiles with energies in the range of $10-25$ MeV/u on targets with mass $A=48-208$.  The case of  $^{48}$Ca(d,p)$^{49}$Ca transfer to the ground state is described in detail.  A comparative study of the effect of the size of experimental errors is also performed.  Five transfer reactions are studied, and their results compiled in order to systematically identify trends.
\item[Conclusions:]  Uncertainties in transfer cross sections can vary significantly (25-100\%) depending on the reaction.  
While these uncertainties are reduced when smaller experimental error bars are used to constrain the potentials, this reduction is not trivially related to the error reduction.
We also find smaller uncertainties when using the adiabatic formulation than when using distorted-wave Born approximation.  
\end{description}
\end{abstract}



\maketitle

\section{Introduction}


The overarching questions in nuclear physics span a wide variety of topics including understanding how visible matter came into being and how it evolves, how subatomic matter organizes itself and what phenomena emerge from this organization, and whether or not the fundamental interactions that govern these structures and evolutions are fully known \cite{LRP2015}.  Many quantities of interest to these goals can be extracted from experiment, but this extraction often relies on reaction theory.  Thus, having a solid understanding of the reaction theory, including the associated uncertainties, is crucial to properly describe the results of these experiments.  There are several sources for the uncertainties in reaction theory  \cite{Lovell2015}: the approximations made in solving the few-body scattering problem, the use of effective interactions and structure functions, a consequence of mapping the many-body problem into a few-body, and finally the influence of the degrees of freedom left out of the model space.   Systematic methods for quantifying the uncertainty introduced by these various sources are needed to move the field forward.

Recently, we used a statistical method to quantify the uncertainty introduced by the effective interactions within a given reaction model \cite{Lovell2017}.  Neutron and deuteron optical potentials were fit to elastic-scattering data through $\chi^2$ minimization, using both an uncorrelated and a correlated $\chi^2$ function. Exploring the $\chi^2$ function around the minimum, we  constructed 95\% confidence bands for the (n,n) and (d,d) elastic cross sections and made predictions for the corresponding (n,n$^\prime$) inelastic and (d,p) transfer cross sections.  This work was done within the frequentist approach.

There are many reasons to pursue uncertainty quantification with Bayesian statistics rather than the frequentist approach.
With Bayesian statistics, we can systematically introduce our prior knowledge into the formulation, and two different theories can be compared and even mixed to provide a better prediction. Moreover, instead of being able to only answer questions where there is a choice of solutions (out of a list of options, one of them must occur - this is the basis of frequentist statistical methods), Bayesian statistics can give probabilities to unique occurrences (such as will it rain tomorrow) \cite{Trotta2008}.  This  provides a more consistent interpretation based on a single set of data, instead of needing multiple occurrences to interpret results in terms of a probability or confidence level.

Although Bayesian methods have been around for centuries, introduced in an essay from the mid-eighteenth century \cite{Bayes1763}, it was only within the past few decades, with the advent of modern Monte Carlo methods, that they became more widely implemented.  In the last several years, nuclear theory has embraced them.  In Effective Field Theories (EFTs), these methods have been particularly effective across a range of applications - parameter estimations \cite{Perez2015,Wesolowski2016,Schindler2009}, assessing uncertainties from model truncations \cite{Furnstahl2015}, understanding how well assumptions hold \cite{Melendez2017}, and  through all of this, propagating uncertainties to observables.  Bayesian methods are also beginning to be widely used in areas such as heavy ion collisions \cite{Bernhard2016, Sangaline2016} and density function theory \cite{Schunck2015,Utama2016}.  

Although all of these references show the power of Bayesian techniques in nuclear theory, there are aspects specific to few-body reaction theory that need to be studied in order to guarantee its usefulness in this subfield.   This work takes the very first step in this direction. Our goals are: i) computing uncertainties on nucleon and deuteron optical model parameters constrained by elastic-scattering data using Bayes' Theorem, ii) determining uncertainties in the  elastic and  (d,p) (or (d,n)) transfer angular distributions, iii) rigorously comparing two approximations in the  theory for transfer reactions, and iv) understanding the information gain on transfer observables, with improvements on elastic-scattering measurements. 

These points are addressed in the following work.  In Section \ref{theory}, we discuss the theoretical framework for this article, including the Bayesian methods used and the reaction formalism for which it is implemented.  The systems that were studied are listed in Section \ref{numbers}, along with some numerical details that are necessary for the work.  Section \ref{results} presents the results from this study, using one detailed case as an example and then summarizing the remaining results, which are discussed in Section \ref{discussion}.  Finally, we conclude in Section \ref{conclusions} with an outlook of how these methods can be further improved.


\section{Theoretical framework}
\label{theory}


\subsection{Bayes' Theorem}

Bayesian methods have become very popular recently in nuclear theory, both because of their power and simplicity (see \cite{Trotta2008,BayesBook} for good introductions). Here we provide a brief summary of the main concepts surrounding Bayes' Theorem so the work is self-contained, and the nonexpert can follow without needing additional background reading.  The main idea that Bayes uses is the fact that the probability of picking two items from a group does not depend on whether you pick item 1 first and then item 2 or item 2 first then item 1: $p(2|1)p(1) = p(1|2)p(2)$, with $p(1)$ and $p(2)$ being the independent probabilities  of picking either item 1 or 2 and $p(2|1)$ being the conditional probability of picking $2$ after having first picked $1$ (viceversa for $p(1|2)$). 

When applied to our field, typically we have an hypothesis $H$ (given by a model) and some constraining external information $D$ (the data). Bayes' theorem is then written as:
\begin{equation}
\label{eqn:bayes}
p(H|D) = \frac{p(H)p(D|H)}{p(D)},
\end{equation}
and provides a method to calculate the posterior distribution, $p(H|D)$, of the hypothesis $H$, conditional on a set of data $D$.  This gives the most likely distribution of the parameters dependent on the given data.  Translating for the application here, the hypothesis  is the optical model for scattering, which introduces many optical potential parameters, and the data is elastic-scattering angular distributions. The question we will try to answer is: what is the most likely distribution for the optical potential parameters given the elastic-scattering data. 

To calculate the posterior distributions with Eq. (\ref{eqn:bayes}), several pieces are needed.  The first is the prior distribution, $p(H)$ (the probability distributions over the various parameters in the optical model), which summarizes our knowledge before the data is seen.  The likelihood function, $p(D|H)$, folds in information about how well the model reproduces the data, typically through a $\chi^2$ function.  In this work, we stick to the standard normal distribution, $e^{-\chi^{2}/2}$, for the likelihood, and the standard definition of the $\chi^2$ distribution,
\begin{equation}
\chi^2 = \frac{1}{N} \sum \limits _{i=1}^N \frac{(\sigma^\mathrm{th}-\sigma^\mathrm{exp})^2}{\Delta \sigma ^2}.
\end{equation}

\noindent The denominator in Eq. (\ref{eqn:bayes}) is the Bayesian evidence, $p(D)$, which typically contains a sum over all possible hypotheses each weighted by their own likelihood function. 

In Bayesian statistics,  95\% confidence intervals are defined by the smallest interval $[a,b]$ for which
\begin{equation}
\label{eqn:CIdef}
\int \limits _a ^b p(H_i|D) dx_i = 0.95,
\end{equation}
\noindent for a given quantity of interest $x_i$.  In practice, for our numerically drawn posteriors, the integral becomes a sum.

For example, to calculate the 95\% confidence intervals for the prior distributions of the constrained and predicted cross sections, at each angle that the calculation was performed, we find the smallest range of cross section values that contains 95\% of the posterior draws.  The minimum and maximum cross section values at each angle define the upper and lower bounds of the 95\% confidence interval.  These intervals then represent the belief that the real value of the cross section of interest has a 95\% chance of falling within that region.  This can be equally calculated for each parameter posterior, which would then define a distribution of the values that the given parameter is most likely to take.

\subsection{Markov Chain Monte Carlo}

While Bayes' theorem is simple in principle, in practice, there can be added complications.  In many cases, calculating the Bayesian evidence numerically is either computationally intractable or impossible.  Then, it is necessary to sample the posterior distribution through a Monte Carlo method.  The longer the sampling, the closer the pulled distribution comes to reproducing the exact posterior distribution.  For this work, we use a Metropolis-Hastings Markov Chain Monte Carlo (MCMC) \cite{Metropolis1953,Hastings1970}.  Because we are interested in calculating free parameters in our model, our hypothesis is a set of parameters, $\boldsymbol{x}$, that define the interaction between projectile and target nuclei.  At each step of the MCMC method, we obtain a set of parameters, $\boldsymbol{x}_i$, from which the prior, $p(H_i)$, and likelihood, $p(D|H_i)$ for the next iteration are obtained.  A new set of parameters, $\boldsymbol{x}_f$, are randomly chosen such that $\boldsymbol{x}_f \sim \mathcal{N}(\boldsymbol{x}_i,\epsilon \boldsymbol{x}_0)$ where $\mathcal{N}$ represents the normal distribution, $\epsilon$ is a scaling factor which controls the step size in parameter space, and $\boldsymbol{x}_0$ is some fixed set of initial parameters; the prior, $p(H_f)$, and likelihood, $p(D|H_f)$, are also calculated with this second set of parameters.  If the condition
\begin{equation}
\label{eqn:MCMC}
\frac{p(H_f)p(D|H_f)}{p(H_i)p(D|H_i)} > R
\end{equation}

\noindent (where $R$ is a random number between 0 and 1) is satisfied, the new parameter set is accepted and becomes the initial parameter set of the next iteration.  If Eq. (\ref{eqn:MCMC}) is not fulfilled, the parameter set is rejected, and a new random set of parameters is drawn from $\mathcal{N}(\boldsymbol{x}_i,\epsilon \boldsymbol{x}_0)$.  This process is continued until a predefined number of parameter sets is accepted.  

There is no guarantee that the initial parameters are within the posterior distribution that we are interested in sampling.  For this reason it is important to have a burn-in process, by rejecting a number of the initial sets obtained with the condition Eq. (\ref{eqn:MCMC}), $N_{burn-in}$ \cite{MCMCBook}. Signatures of a good burn-in and healthy sampling are likelihoods and parameter distributions that oscillate around a mean. Following the burn-in, each accepted parameter set is dependent on the previously accepted set. To remove this dependence, one needs to reject $N_{jump}$ sets in between each accepted set, so erroneous correlations are not introduced.  More details on implementing MCMC can be found in \cite{MCMCBook,RobertCasella2004}.

\subsection{Optical model}

In this work, we constrain the parameters within the optical model, which describe the scattering of  a projectile on a target by solving the single-channel scattering equation, in the center of mass system, with an effective interaction $U(r)$, $r$ being the relative coordinate between projectile and target.  The so-called optical potentials are characterized by both real and imaginary terms,
\begin{equation}
U(r) = V(r) + i W(r).
\end{equation}

\noindent The imaginary term takes into account flux that leaves the elastic channel and is not explicitly described by the reaction model.  

These potentials generally have volume and surface parts which are written as Woods-Saxon shapes or derivatives of Woods-Saxon shapes.  Regardless of whether the real volume term
\begin{equation}
V(r) = -\frac{V_V}{1+\mathrm{exp}(\frac{r-R_V}{a_V})},
\end{equation}

\noindent or imaginary volume term
\begin{equation}
W(r) = -\frac{W_V}{1+\mathrm{exp}(\frac{r-R_W}{a_W})},
\end{equation}

\noindent is considered, each term contains three free parameters, a depth, radius  and diffuseness. The radii of the optical potential terms are often parameterized in terms of a radius parameter $r_i$, and for the cases we here consider,  $R_i=r_iA^{1/3}$ with $A$ being the mass number of the target. The surface term is typically purely imaginary and written as the derivative of a Woods-Saxon.  These three terms, real volume, imaginary volume, and imaginary surface, introduce 9 parameters.

In additional to the nuclear central potential, there is a spin-orbit term, typically parameterized by a Woods-Saxon derivative and, for charged projectiles, a Coulomb potential. Because the spin-orbit does not strongly influence the elastic scattering cross section for nucleons and deuterons, in this work, we fix the parameters for the spin-orbit term at chosen initial values. We consider the standard finite-size Coulomb potential, (e.g. \cite{Fukui2014}), parametrized by a Coulomb radius, which we also keep fixed throughout this work.

\subsection{Describing transfer reactions}

There are two reaction models that we consider.  The first is the adiabatic wave approximation (ADWA) which starts from a three-body description of $n+p+A$ and relies on the separation between a fast and a slow variable, namely the fast center of mass of a projectile-target system, $\vec{R}$, compared to the slow internal motion of the deuteron, $\vec{r}$ \cite{ReactionsBook}(Chapter 7.1).  This approximation consists of exactly solving
\begin{equation}
[T_R + V_{pA} + V_{nA} -(E-\epsilon_0)]\Psi ^{\mathrm{ad}} (\vec{r},\vec{R}) = 0,
\label{eqn:adwa}
\end{equation}
\noindent where $T_R$ is the kinetic energy of the center of mass, $V_{pA}$ and $V_{nA}$ are the optical potentials that describe the proton-target and neutron-target interactions, and $E$ is the incoming beam energy. The term $\epsilon_0$ is the ground state energy of the deuteron and arises from the adiabatic approximation where the deuteron breakup states can be made degenerate with the ground state. In the adiabatic method, breakup of the deuteron is treated to all orders \cite{Johnson1974}. A discussion of breakup and finite-range effects in ADWA can be found in \cite{Nguyen2010}.   
The adiabatic wave function is then introduced in the post-form T-matrix for the A(d,p)B transfer process \cite{ReactionsBook}:
\begin{equation}
\boldsymbol{T}^{\mathrm{ADWA}}_{\mathrm{post}} = \langle \Phi_{nA} (\vec{r}_{nA}) \chi_p (\vec{R}_f) | V_{np} | \Psi ^{\mathrm{ad}} (\vec{r},\vec{R}) \rangle ,
\end{equation}
where $\Phi _{nA}$ is the bound-state wave function of $B=n+A$, $\chi_p$ is the outgoing proton wave function, $V_{np}$ is the deuteron binding potential, and the remnant term $\Delta$, corresponding to the difference between the $A+p$ and $B+p$ optical potentials, is neglected. From the T-matrix, angular distributions for the (d,p) cross sections can be readily obtained \cite{ReactionsBook}.

\vspace{0.5cm}

A simpler approach is the distorted wave Born approximation (DWBA),  often used when interpreting $A(d,p)B$ data. This is a perturbative approach,  which, when taken to first order, assumes the reaction takes place in one-step and only includes breakup effectively through the elastic deuteron channel. Then, the elastic scattering of the deuteron is described by an effective deuteron-target interaction, $U_{dA}$ (as opposed to ADWA that uses the individual nucleon-target interactions).  Solving the single-channel scattering equation with $U_{dA}$ provides the  distorted wave $\chi_d$ for the deuteron relative to the target.  The post-form DWBA T-matrix for the A(d,p)B is given by \cite{ReactionsBook}:
\begin{equation}
\boldsymbol{T}^{\mathrm{DWBA}}_{\mathrm{post}} = \langle \Phi_{nA} (\vec{r}_{nA}) \chi_p (\vec{R}_f) | V_{np}| \Phi _{np}(\vec{r}_{np}) \chi _{d\vec{k}_{i}}(\vec{R}_i) \rangle,
\end{equation}
where the initial three-body state is replaced by the elastic deuteron channel, namely the product of the initial deuteron bound state, $\Phi _{np}(\vec{r}_{np})$, and the distorted wave of the deuteron relative to the target $\chi _{d\vec{k}_{i}}(\vec{R}_i)$. \\

In the equations for ADWA and DWBA presented above we have assumed (d,p) reactions. These can be trivially reformulated for (d,n) reactions.  In this case, $B = A+p$, and $\chi_p$ is replaced by the distorted wave of the outgoing $n+B$ system.

\section{Numerical Details}
\label{numbers}


For this work we studied five transfer reactions using both ADWA and DWBA as described in Section \ref{theory}:  $^{48}$Ca(d,p)$^{49}$Ca at 24 MeV, $^{90}$Zr(d,p)$^{91}$Zr at 22 MeV, $^{90}$Zr(d,n)$^{91}$Nb at  20 MeV, $^{116}$Sn(d,p)$^{117}$Sn at 44 MeV and $^{208}$Pb(d,p)$^{209}$Pb at 32 MeV.
These were chosen specifically because, in addition to the transfer data, there were all relevant elastic-scattering data to constrain the optical potentials both in the entrance and exit channel.
 Table \ref{tab:data} gives a list of the relevant elastic-scattering reactions, along with the reference to the experimental data.  The starting potentials for the nucleon elastic scattering was taken from Becchetti and Greenlees \cite{BGOpt}, and the starting potentials for the deuteron elastic scattering were from An and Cai \cite{ACOpt}.  

\begin{table}
\begin{center}
\begin{tabular}{cccc}
\hline \hline \textbf{Target} & \textbf{Projectile} & \textbf{E (MeV)} & \textbf{Data} \\ \hline
$^{48}$Ca & $p$ & 14.03 & \cite{48Capp1403} \\ 
$^{48}$Ca & $n$ & 12.0 & \cite{48Cann12} \\
$^{48}$Ca & $p$ & 24.0 & \cite{48Capp25} \\
$^{48}$Ca & $d$ & 23.3 & \cite{48Ca90Zrdd232} \\
$^{90}$Zr & $p$ & 12.7 & \cite{90Zrpp127} \\
$^{90}$Zr & $n$ & 10.0 & \cite{90Zrnn10nn24} \\
$^{90}$Zr & $p$ & 22.5 & \cite{90Zrpp225} \\
$^{90}$Zr & $n$ & 24.0 & \cite{90Zrnn10nn24} \\
$^{90}$Zr & $d$ & 23.2 & \cite{48Ca90Zrdd232} \\
$^{116}$Sn & $p$ & 22.0 & \cite{116Snpp22} \\
$^{116}$Sn & $n$ & 24.0 & \cite{116Snnn24} \\
$^{116}$Sn & $p$ & 49.35 & \cite{116Snpp4935} \\
$^{208}$Pb & $p$ & 16.9 & \cite{208Pbpp169} \\
$^{208}$Pb & $n$ & 16.0 & \cite{208Pbnn16} \\
$^{208}$Pb & $p$ & 35.0 & \cite{208Pbpp35} \\
$^{208}$Pb & $d$ & 28.8 & \cite{208Pbdd288} \\ \hline \hline
\end{tabular}
\caption{Summary of elastic-scattering pairs used in this work.  Column four gives the corresponding reference for the experimental data.}
\label{tab:data}
\end{center}
\end{table}

For the ADWA calculations, we study the individual and combined uncertainties from the incoming proton- and neutron-target interactions as well as from the outgoing nucleon-(A+1) interaction.  Likewise, for the DWBA calculations, we study the uncertainties coming from the deuteron-target interaction and outgoing nucleon-(A+1) interaction.  

The interaction  between the neutron and proton of the deuteron is taken to have a Gaussian shape which reproduces the nucleon separation energy of the system.  The interaction describing the final bound state between the transferred nucleon and the target is taken to be a central potential (Woods-Saxon shape) with a spin-orbit term.  These two terms are parameterized with typical radius and diffuseness values of $1.20$ fm and $0.65$ fm.  The depth of the spin-orbit term is also set to a typical value of $6.0$ MeV, while the depth of the central term is fit to reproduce the binding energy of the nucleon-target system.  

For the majority of this work, we consider a Gaussian prior.  An individual Gaussian is defined for each optical model parameter, centered on the original parameter value (from either \cite{BGOpt} or \cite{ACOpt}) and with a width of the original parameter value.  This is discussed in more detail in Section \ref{sec:priors}.  

As mentioned in Section \ref{theory}, the MCMC method requires some specifications that must be adjusted for each system (burn-in, step size, etc.).  The step taken by each parameter for the Monte Carlo in this work is drawn from a Gaussian distribution, with a width defined to be a percentage of the starting parameter value ($\epsilon x_0^i$).  In this way, the step size is not dependent on the previous parameter draw and has the appropriate scale.  Throughout this work, $\epsilon = 0.005$, which results in a 50\% rejection of the tested parameters by the Monte Carlo method, as typical in other applications \cite{MCMCBook}.

\begin{figure}[h!]
\begin{center}
\epsfig{file=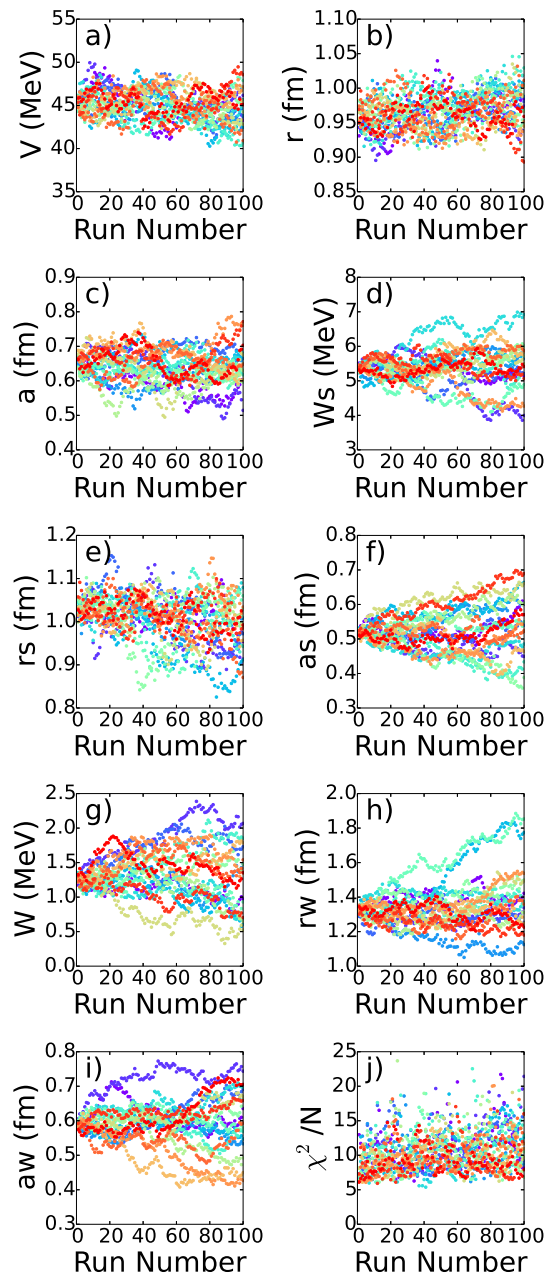,width=0.4\textwidth} 
\caption{(Color online) Representation of the the MCMC samplings for $^{90}$Zr(n,n)$^{90}$Zr at 24.0 MeV.  Different shades of gray (different colors online) represent different processors.}
\label{fig:MCMCresults}
\end{center}
\end{figure}

In our work we use $N_{burn-in}=500$, which is  sufficient to get close to a minimum in parameter space constrained by the prior distribution.  By examining the $\chi^2$ and likelihood values recorded after the burn-in, we can verify that these values are no longer changing systematically and that the posterior distribution is being adequately sampled.  Also, to insure that each of the parameter set is pulled independently from the previous one, we set $N_{jump}=10$ and record one out of every ten accepted parameter sets.  A representation of a set of draws is shown in Figure \ref{fig:MCMCresults}.

For many of the experimental data sets that we use, the quoted errors are due to the digitalization of the data from the original publications and do not reflect the actual experimental error.  When errors are quoted in the original work, they often only include statistical errors and not systematic ones, which tend to be larger.  For these reasons, we initially take all experimental errors to be 10\% of the experimental data.  Then, we can also systematically study the effect that the reduction of experimental error bars has on the overall transfer cross sections and any extracted quantities.

The Monte Carlo methods and Bayesian analysis that is discussed here is newly implemented but it makes use of the codes {\sc fresco} \cite{fresco} to calculate elastic cross sections, {\sc sfresco} \cite{fresco} to constrain the cross sections based on data (using the {\sc minuit} \cite{minuit} minimization routine), and {\sc nlat} \cite{NLAT} to calculate transfer cross sections in ADWA and DWBA.


\section{Results}
\label{results}

\subsection{Dependence on the prior}
\label{sec:priors}



In Bayesian statistics, there is an interplay between the prior distributions and the likelihood (given by the data) to produce the posteriors. In principle, well-measured data will cause the likelihood to dominate over the prior distribution,  negating the influence of the prior \cite{Trotta2008}.  In order to understand the relevance of the prior in these sort of reactions, it is important to investigate the effect it has on our parameter posterior distributions and on the resulting cross-section confidence intervals.  

\begin{figure}[h!]
\begin{center}
\epsfig{file=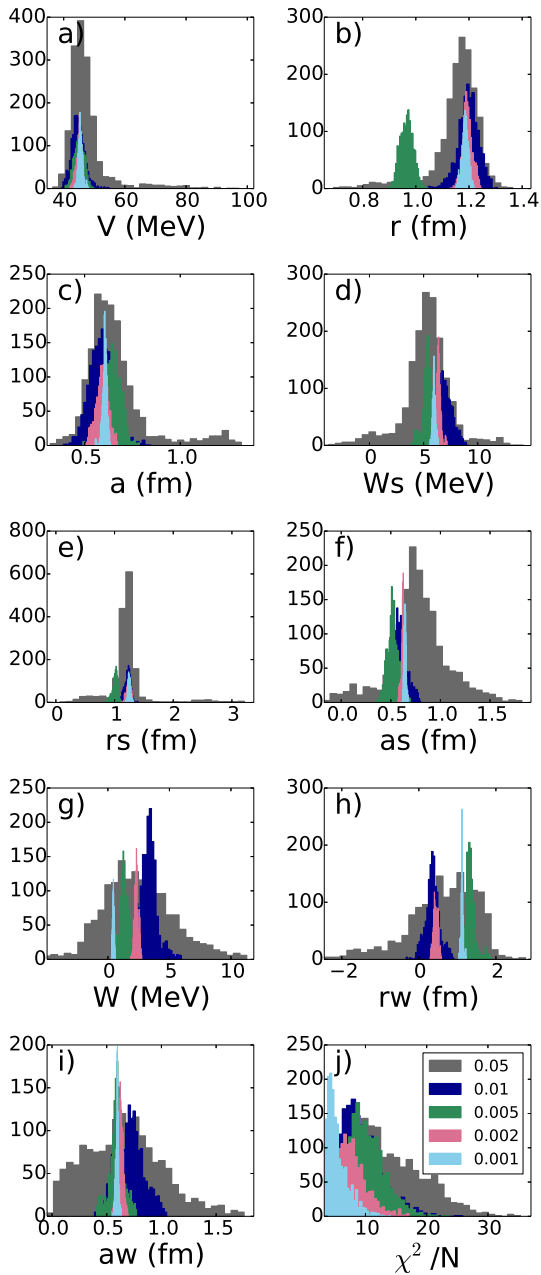,width=0.4\textwidth}
\caption{(Color online)  Comparison of the posterior distributions (histograms) resulting from various prior distributions (corresponding solid lines) for a wide Gaussian (WG), medium Gaussian (MG), and narrow Gaussian (NG) as defined in Table \ref{tab:priorshape} for $^{90}$Zr(n,n)$^{90}$Zr at 24.0 MeV.}
\label{fig:testprior}
\end{center}
\end{figure}

\begin{figure}
\begin{center}
\includegraphics[width=0.4\textwidth]{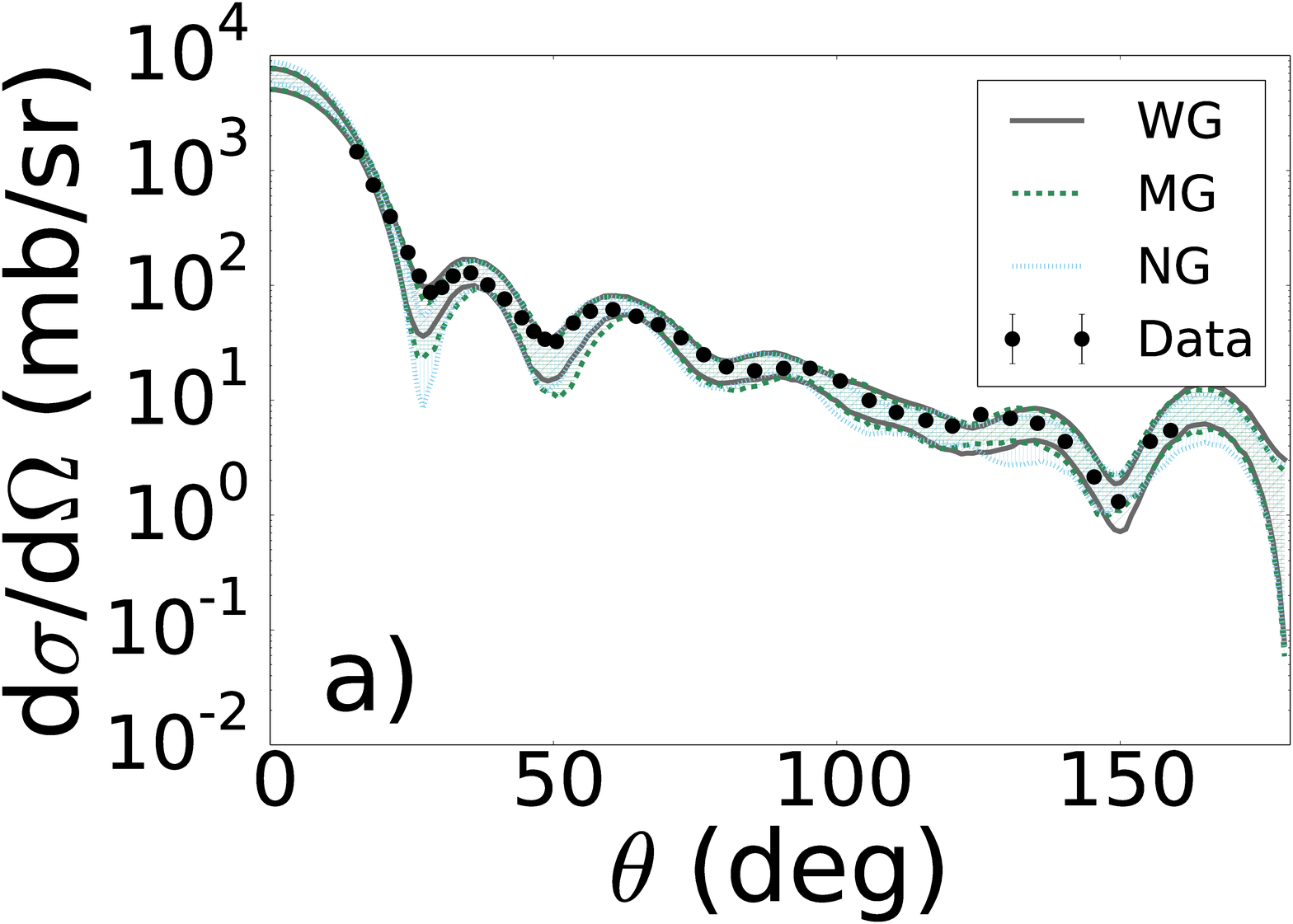} \\
\includegraphics[width=0.4\textwidth]{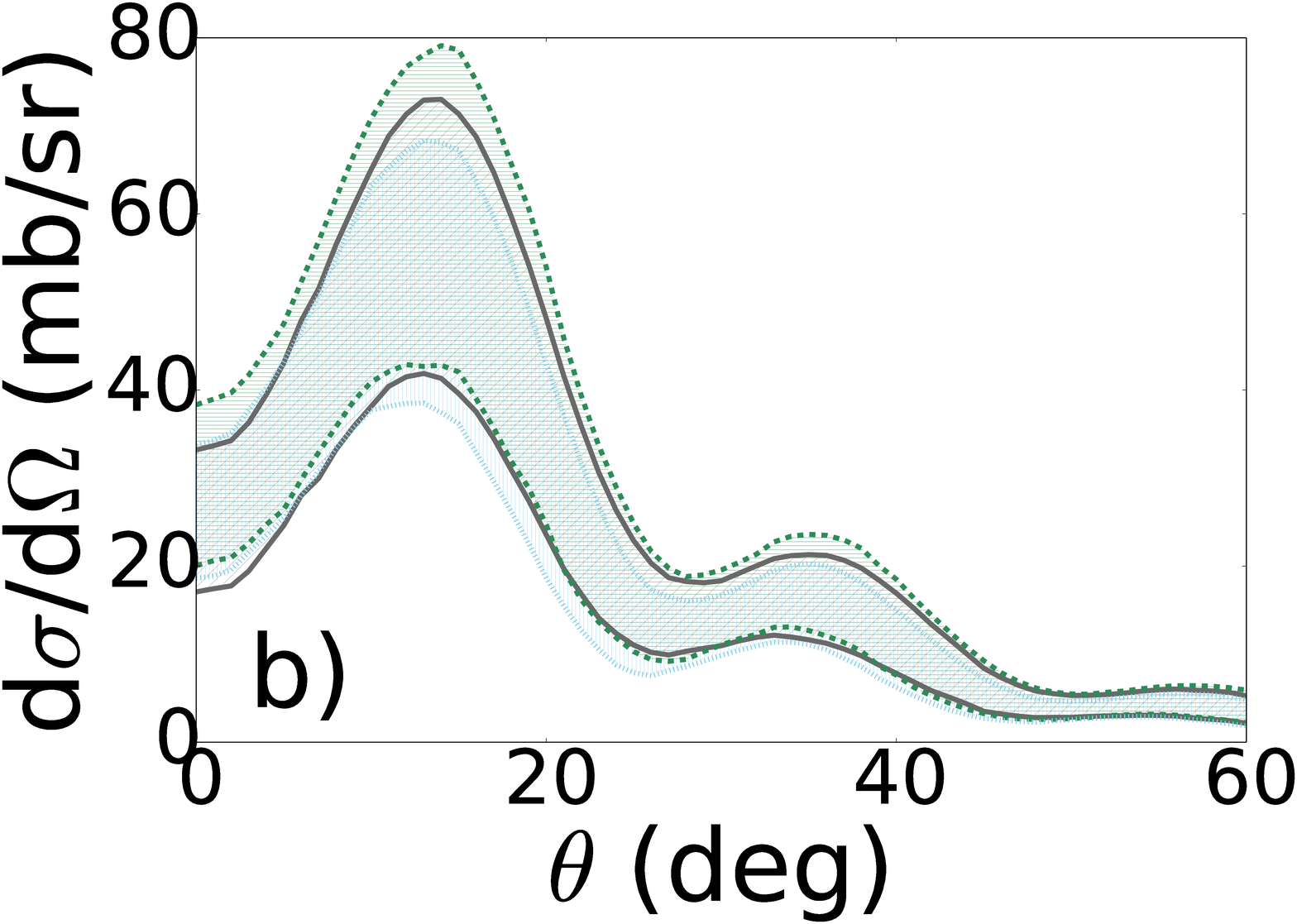}
\caption{(Color online)  Comparison of the (a) $^{90}$Zr(n,n)$^{90}$Zr elastic scattering and (b) $^{90}$Zr(d,p)$^{91}$Zr(g.s.) transfer at 24.0 MeV for the posterior distributions shown in Figure \ref{fig:testprior}.}
\label{fig:90Zrn24elasticprior}
\end{center}
\end{figure}

\begin{figure}
\begin{center}
\includegraphics[width=0.4\textwidth]{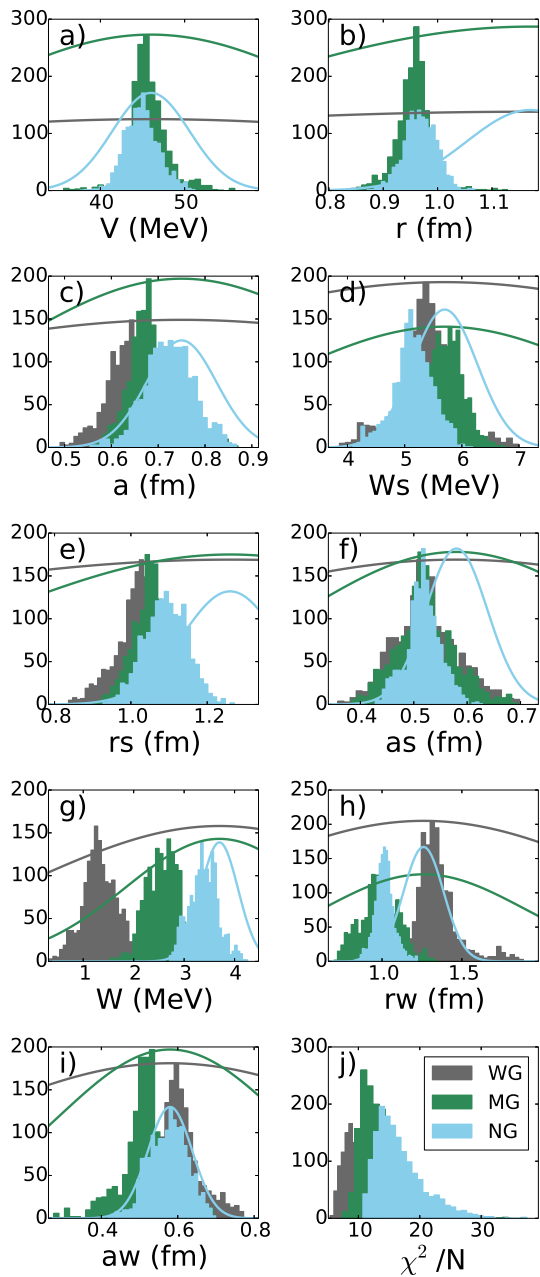} 
\caption{(Color online)  Same as Figure \ref{fig:testprior} for a fixed prior shape - wide Gaussian - but varying $\epsilon$, as given in the legend in panel (j).}
\label{fig:teststep}
\end{center}
\end{figure}

Before focusing on our physics cases, we compared prior distributions for four elastic scattering data sets:  $^{48}$Ca(p,p) at 21.0 MeV, $^{90}$Zr(p,p) at 40.0 MeV, $^{120}$Sn(n,n) at 13.9 MeV, and $^{90}$Zr(n,n) at 24.0 MeV, but all results shown in Figures \ref{fig:testprior}, \ref{fig:90Zrn24elasticprior}, \ref{fig:teststep}, and \ref{fig:syserrors} pertain to $^{90}$Zr(n,n).  These covere a range of masses and energies, as well as include both neutron and proton projectiles.  We tested both a linear (flat) prior and a Gaussian prior, each one with a wide width (covering a range of parameter space much larger than the expected physical range of the parameters) and a medium width (covering the expected physical range of the parameter space).  The centers and widths of these priors are listed in Table \ref{tab:priorshape} specified for $^{90}$Zr(n,n)$^{90}$Zr at 24.0 MeV.  For this case, we also compare these four priors to narrow linear and Gaussian priors to illustrate the effect of stringent prior limits on the potential and resulting observables.  The narrow priors have widths of 10\% the initial value for each parameter, medium priors have widths of 50\%, and wide priors have widths of 100\%.


\begin{table*}
\begin{center}
\begin{tabular}{c|cccc|cc|cc|cc|cc|cc|cc}
\hline \hline 
\textbf{$x$} & \textbf{$x_o$} & \textbf{$\Delta x_{\mathrm{N}}$} & \textbf{$\Delta x _{\mathrm{M}}$} & \textbf{$\Delta x_{\mathrm{W}}$} & \textbf{$x_{oWL}$} & \textbf{$\Delta x_{WL}$} & \textbf{$x_{oML}$} & \textbf{$\Delta x_{ML}$} & \textbf{$x_{oNL}$} & \textbf{$\Delta x_{NL}$} & \textbf{$x_{oWG}$} & \textbf{$\Delta x_{WG}$} & \textbf{$x_{oMG}$} & \textbf{$\Delta x_{MG}$} & \textbf{$x_{oNG}$} & \textbf{$\Delta x_{NG}$} \\ \hline
$V$ & 46.0 & 4.6 & 23.0 & 46.0 & 45.86 & 2.45 & 43.69 & 2.08  & 45.95 & 1.28 & 45.06 & 1.68 & 45.68 & 2.29 & 44.92 & 2.01 \\
$r$ & 1.17 & 0.117 & 0.585 & 1.17 & 0.95 & 0.03 & 0.98 & 0.03 & 1.18 & 0.02 & 0.97 & 0.02 & 0.95 & 0.03 & 0.97 & 0.03 \\
$a$ & 0.75 & 0.075 & 0.375 & 0.75 & 0.69 & 0.06 & 0.63 &0.04 & 0.77 & 0.02 & 0.64& 0.05 & 0.68 & 0.05 & 0.72 & 0.05 \\
$W_s$ & 5.7 & 0.57 & 2.85 & 5.7 & 4.76 & 0.55 & 4.95 & 0.47 & 5.70 & 0.16 & 5.39 & 0.50 & 4.70 & 0.32 & 5.15 & 0.35 \\
$r_s$ & 1.26 & 0.126 & 0.63 & 1.26 & 1.06 & 0.07 & 1.03 & 0.05 & 1.27 & 0.03 & 1.01 & 0.05 & 1.05 & 0.05 & 1.09 & 0.06 \\
$a_s$ & 0.58 & 0.058 & 0.29 & 0.58 & 0.61 & 0.04 & 0.48 & 0.03 & 0.58 & 0.02 & 0.52 & 0.06 & 0.52 & 0.05 & 0.52 & 0.03 \\
$W$ & 3.7 & 0.37 & 1.85 & 3.7 & 2.15 & 0.22 & 3.00 & 0.17 & 3.66 & 0.10 & 1.34 & 0.35 & 2.64 & 0.40 & 3.39 & 0.28 \\
$r_w$ & 1.26 & 0.126 & 0.63 & 1.26 & 1.06 & 0.10 & 1.10 & 0.07 & 1.25 & 0.03 & 1.35 & 0.13 & 0.99 & 0.12 & 1.02 & 0.06 \\
$a_w$ & 0.58 & 0.058 & 0.29 & 0.58 & 0.57 & 0.07 & 0.68 & 0.04 & 0.58 & 0.02 & 0.59 & 0.07 & 0.52 & 0.07 & 0.57 & 0.05 \\ \hline \hline
\end{tabular}
\caption{Summary of centers ($x_o$) and widths ($\Delta x_N$, $\Delta x_M$, and $\Delta x_W$) for the narrow, medium, and wide priors, and the resulting means and widths for the narrow linear (NL), narrow Gaussian (NG), medium linear (ML), medium Gaussian (MG), wide linear (WL), and wide Gaussian (WG) posterior distributions.  These are given for $^{90}$Zr(n,n)$^{90}$Zr at 24.0 MeV.  Depths are given in MeV, and radii and diffusenesses are given in fm.}
\label{tab:priorshape}
\end{center}
\end{table*}

We first fix $\epsilon=0.005$ and compare the posterior distributions for the priors, shown in Figure \ref{fig:testprior}.  Although we studied the six priors shown in Table \ref{tab:priorshape}, we only show the distributions for the Gaussian priors for ease of viewing.  The solid lines show the shape and range of the prior distributions, and the histograms show the resulting posterior distributions.  For the real parameters, especially $V$ and $r$, (panels a and b) the posterior distributions are nearly identical, within statistical fluctuations.  This is not necessarily the case in the imaginary terms of the potential, especially for the volume part.  For the volume depth, $W$ in panel g, the peaks and widths of the three distributions shown are strikingly different.  For the Gaussian priors systematic decreases in the depth lead to increases in $r_w$, panel h.  Even so, all of the parameter values are reasonable.  These differences in the minima lead to similar $\chi^2$ distributions, also seen in Figure \ref{fig:testprior} panel j.  

These conclusions do not necessarily hold for the linear priors (shown in more detail in Appendix \ref{app:priors}).  The posterior distributions for the real volume potential are similar to those resulting from the Gaussian priors, but many of the imaginary parameters have hard boundaries in the posterior distributions due to the sharp cut-offs of the prior distributions.  These parameter space cut-offs can significantly influence the resulting 95\% confidence intervals.  The parameters constrained by the narrow linear prior, for example, do not reproduce the elastic scattering data.  This, in part, leads to our use of Gaussian priors.  


Figure \ref{fig:90Zrn24elasticprior} (a) then shows the comparison of the distribution of elastic cross section values calculated from these parameter posteriors.  The bands are constructed by calculating an angular distribution from each posterior and then computing the 95\% confidence interval for each angle (as in Eq. \ref{eqn:CIdef}).  They are nearly identical, except for minor fluctuations especially at the first minimum and backwards angles.  This is perhaps unsurprising given the similarities of the $\chi^2$ distributions, but it  indicates that parameters are not uniquely determined by the data.  

A given set of optical potentials can produce considerably different transfer cross section,  even when they produce identical elastic scattering distributions (for example, \cite{Lovell2015}).  We thus investigate the transfer cross sections resulting from the posteriors shown in Fig. \ref{fig:testprior}.  Given the computational costs of the full calculation, we use only DWBA and simplify the process by setting $U_{Ad} = 2U_{An}$, since the proton- and neutron-target potentials are rather similar.  Doing this, and defining the the $^{90}\mathrm{Zr}-p$ outgoing channel using \cite{BGOpt}, we calculate the resulting transfer cross sections shown in Figure \ref{fig:90Zrn24elasticprior} (b). The transfer cross sections show more pronounced differences than the elastic.  In particular, the narrow Gaussian prior produces a cross section with a slightly reduced magnitude.  We expect that if the prior distribution is large enough, the effect of the shape of the prior disappears entirely.  

We can then fix the shape of the prior and examine the effect of varying the scale $\epsilon$.  Even though $\epsilon = 0.005$ gives us near the ideal relation between the number of accepted and rejected steps, we aim to verify that this adequately explores the parameter space: too small a step for each parameter can result in trapping the random walk in a local minimum near the initial parameterization, instead of finding a more global minimum, which then can produce an artificially narrow posterior.  Assuming the wide Gaussian prior, we repeated the calculations for $\epsilon = 0.001, 0.002, 0.005, 0.01, 0.05$; the corresponding posterior distributions are shown in Fig. \ref{fig:teststep}. Our results show that, on one hand $\epsilon = 0.05$ (black) is too large and not able to constrain the posterior distribution for the parameters, and on the other, $\epsilon=0.001$ and $\epsilon=0.002$ result in extremely narrow posterior distributions, indicating that these steps do not allow enough exploration of the parameter space. 

For $\epsilon=0.005$ and $\epsilon=0.01$, the right balance is provided:  we obtain nearly the same posterior widths for each parameter, and close to identical $\chi^2$ distributions, as seen in panel (j) (overlapping green and blue histograms). The parameter posterior distributions are reasonable, and although they do result in slightly different mean values for the parameters, the 95\% confidence intervals on the elastic scattering cross sections for these two $\epsilon$ choices are identical. Note that it is the cross section that is the observable, not the potential; the individual parameter posterior distributions are less important than the combined effect of all of the parameters.  Our choice is to use $\epsilon = 0.005$ throughout the rest of this work.

These trends in parameter shape and $\epsilon$ hold for the other three reactions that were studied.  In these cases, the posterior distributions for the flat priors had sharp problematic cut-offs which lead to obvious differences in the calculated cross sections.  For this reason, we discard the flat priors, and only use Gaussian priors for the remainder of this work. 

\subsection{Prior influence on the posterior}

\begin{figure}
\begin{center}
\includegraphics[width=0.4\textwidth]{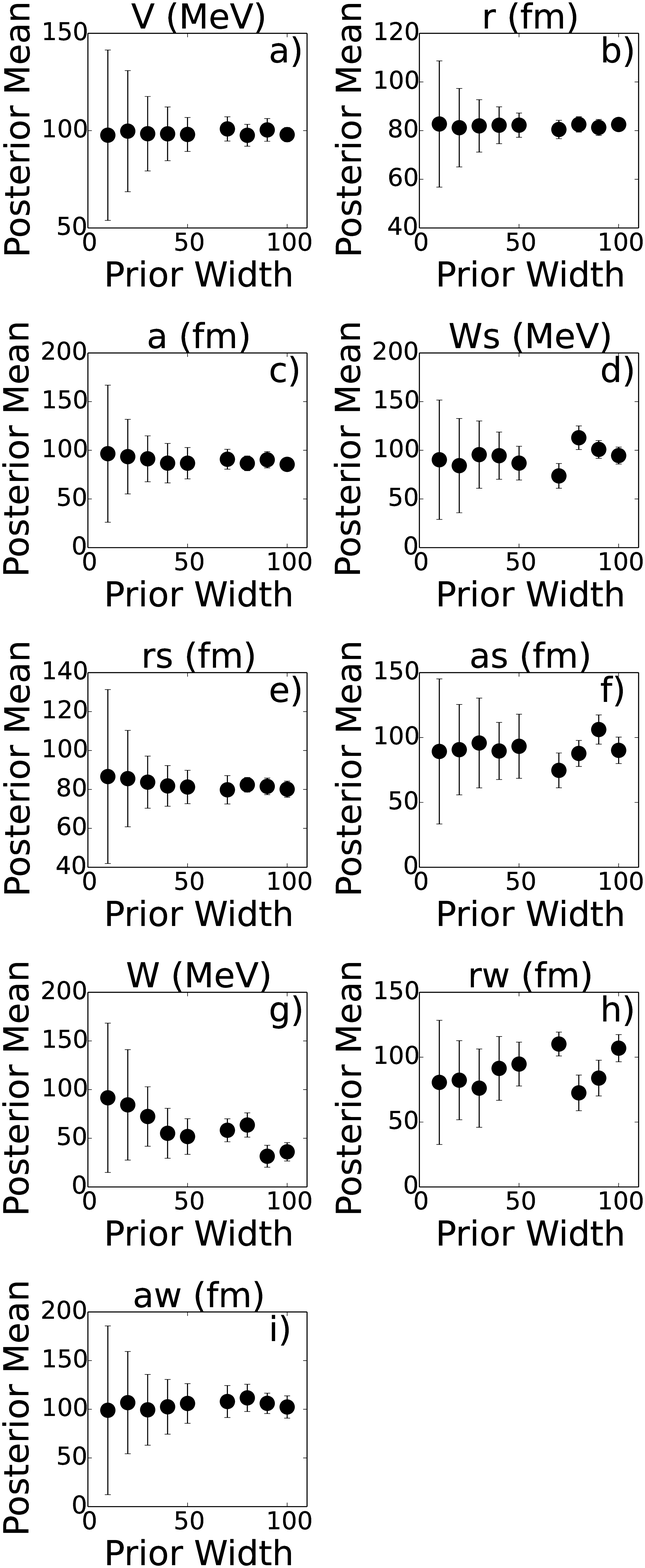} 
\caption{(Color online)  Systematic study of the mean of the posterior distribution as a function of the width of the prior distribution for neutron elastic scattering on $^{90}$Zr at 24.0 MeV:  the prior width along the x-axis is given as the percentage of the original parameter value from \cite{BGOpt}; posterior mean is given as a percentage of the prior center (full circles);  the error bars show the width of the posterior mean as a percentage of the width of the prior.}
\label{fig:syserrors}
\end{center}
\end{figure}

We further systematically study how the mean and width of the posterior change with the width of the prior.  To do this, we again take the example of $^{90}$Zr(n,n)$^{90}$Zr elastic scattering at 24.0 MeV and use a Gaussian prior for each of the parameters, with a mean value of the starting parameter value from \cite{BGOpt} and a varying width - defined as a percentage of the mean value.  Figure \ref{fig:syserrors} shows the means (filled circles) and widths (error bars) of the resulting posterior distributions as a function of the width of the prior distributions for each optical potential parameter.  Clearly, the mean of the posterior distribution for each parameter stays essentially constant as the width of the prior is increased.  

However, the resulting posterior widths are significantly narrower than the starting prior distributions - and this width does not depend on the width of the prior.  For the following results, we take the width of the prior distribution to be the same value as the original parameter ($\Delta x = x_o$) for each of the optical model parameters that are allowed to vary.  This keeps the parameters within a physical range but does not overly constrain them.

\subsection{Transfer reactions}

For the main part of this work, we studied five transfer reactions, four neutron transfers and one proton transfer, as listed in Section \ref{numbers}.  In the next section, we will show the example of $^{48}$Ca(d,p)$^{49}$Ca in detail, first going through the ADWA and then the DWBA calculations.  Following that, we will summarize our findings of all five reactions, discussing specific details as well as systematic trends.


\subsubsection{Transfer using ADWA}
\label{ADWA}


We first calculate $^{48}$Ca(d,p)$^{49}$Ca transfer (to the ground state) using the adiabatic wave approximation (ADWA).  The incoming deuteron channel is constrained using neutron and proton elastic scattering on $^{48}$Ca.  The outgoing proton-$^{49}$Ca channel is constrained using proton scattering data on $^{48}$Ca at an energy in the center of mass that is approximately $2E^{CM}_d - Q_{(d,p)}$.  This is appropriate because, for the nuclei considered here, the differences in the nucleon optical potentials between the $A$ and the $A+1$ systems are typically on the order of 1\% or less.

To complete the transfer reaction, we need to calculate the posterior distributions for each of these elastic scattering cases.  As discussed in the previous section, we use an independent Gaussian prior for each parameter ($x^i$) centered on the starting parameterization ($x_0^i$ for each parameter) from \cite{BGOpt} and the width is the same as the center value,
\begin{equation}
p(H) \propto \prod \limits _{i=1} ^{N_{p}} \mathrm{exp}\left [-\frac{(x^i-x^i_0)^2}{2(x^i_0)^2}\right ],
\label{eqn:Gauprior}
\end{equation}

\noindent where $N_p$ is the number of parameters that are being constrained.

\begin{figure}
\begin{center}
\includegraphics[width=0.4\textwidth]{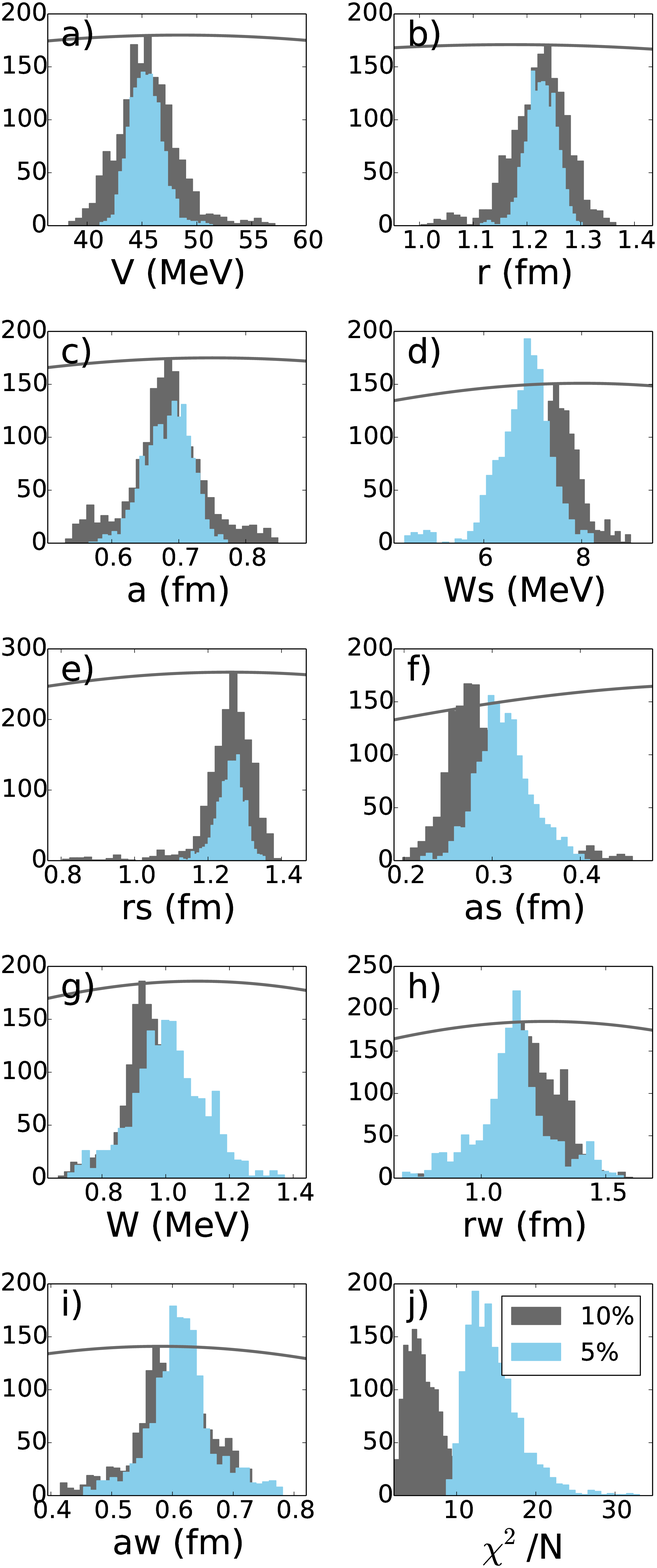}
\caption{(Color online)  Posterior distributions for the optical model parameters conditional on $^{48}$Ca(n,n) elastic scattering at 12.0 MeV.  Gray (light blue) histograms show the posterior assuming 10\% (5\%) error on the experimental data.  Overlaid solid line shows the Gaussian prior distribution (magnitude is arbitrary).}
\label{fig:48Can12}
\end{center}
\end{figure}

\begin{figure}
\begin{center}
\includegraphics[width=0.4\textwidth]{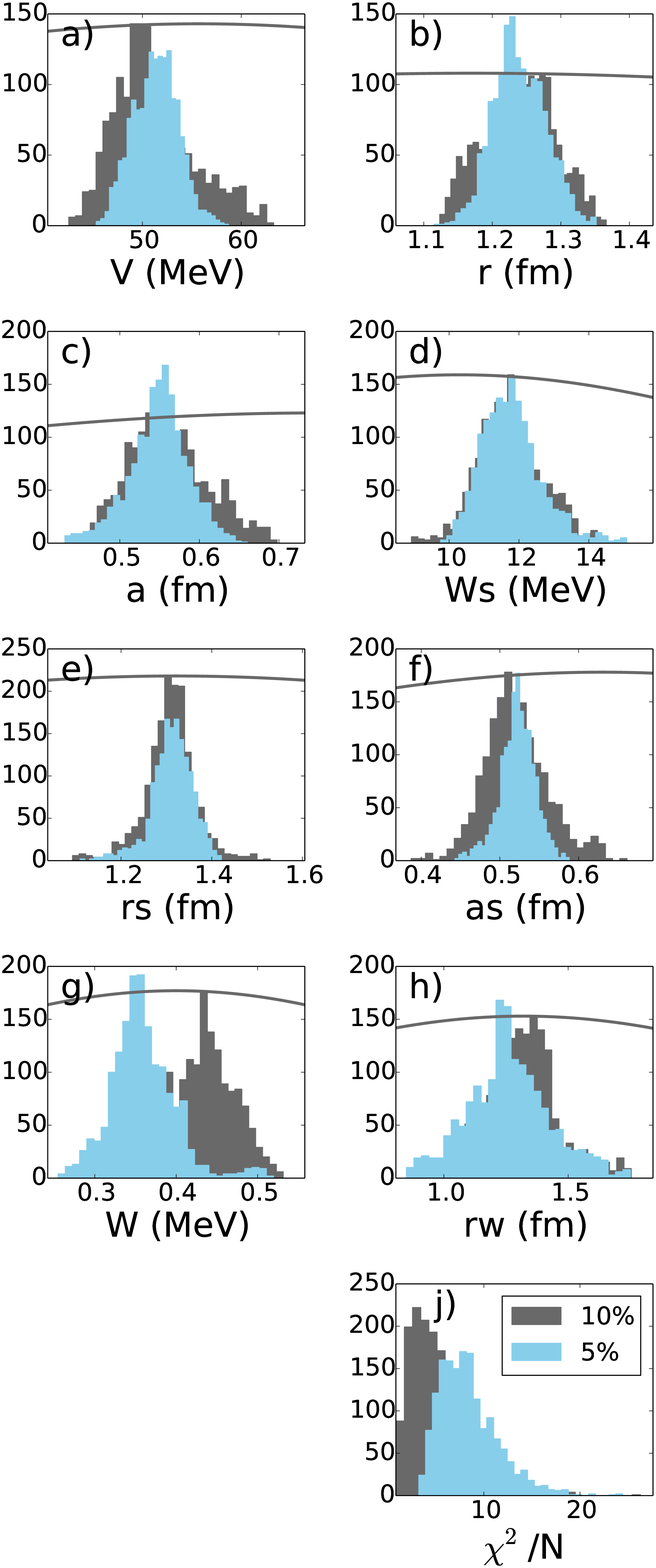} 
\caption{(Color online)  Same as Figure \ref{fig:48Can12} for $^{48}$Ca(p,p) elastic scattering at 14.03 MeV.}
\label{fig:48Cap1403}
\end{center}
\end{figure}

\begin{figure}
\begin{center}
\includegraphics[width=0.4\textwidth]{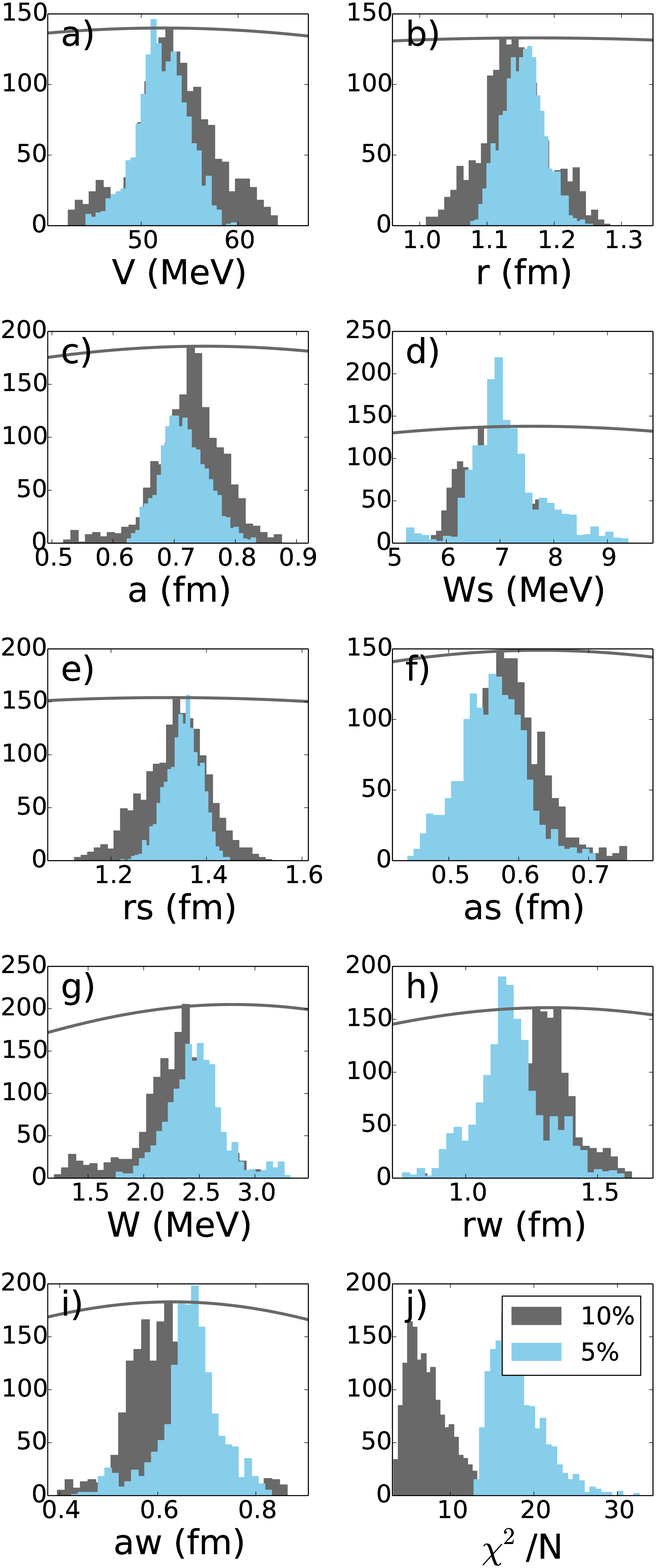}
\caption{(Color online)  Same as Figure \ref{fig:48Can12} for $^{48}$Ca(p,p) elastic scattering at 25.0 MeV.}
\label{fig:48Cap25}
\end{center}
\end{figure}

In Figures \ref{fig:48Can12}, \ref{fig:48Cap1403}, and \ref{fig:48Cap25}, we show the prior (solid line) and posterior (black histogram) distributions for each of the variables that were constrained by data (taking the experimental error bars to be 10\% of the data).  These distributions were constructed from 1600 accepted MCMC draws.  Each of the posterior distributions is centered around a physical value with a width that is significantly narrower than the width of the prior.  This demonstrated that the data has important information content pertaining these parameters.

We note that Figure \ref{fig:48Cap1403} does not show a plot for $a_w$, the imaginary volume diffuseness.  In this case, $a_w$ was not included as a free parameter in the Monte Carlo simulation, but was instead fixed at its initial value of 0.63 fm.  When it was included as a free parameter, it was not well constrained by the data;  the posterior distribution was completely flat and outside of the range defined by the prior (as well as outside of the physical range for this parameter).  This was not uncommon for the proton-scattering and deuteron-scattering data, as will be discussed in Section \ref{discussion}.

\begin{figure}
\begin{center}
\includegraphics[width=0.4\textwidth]{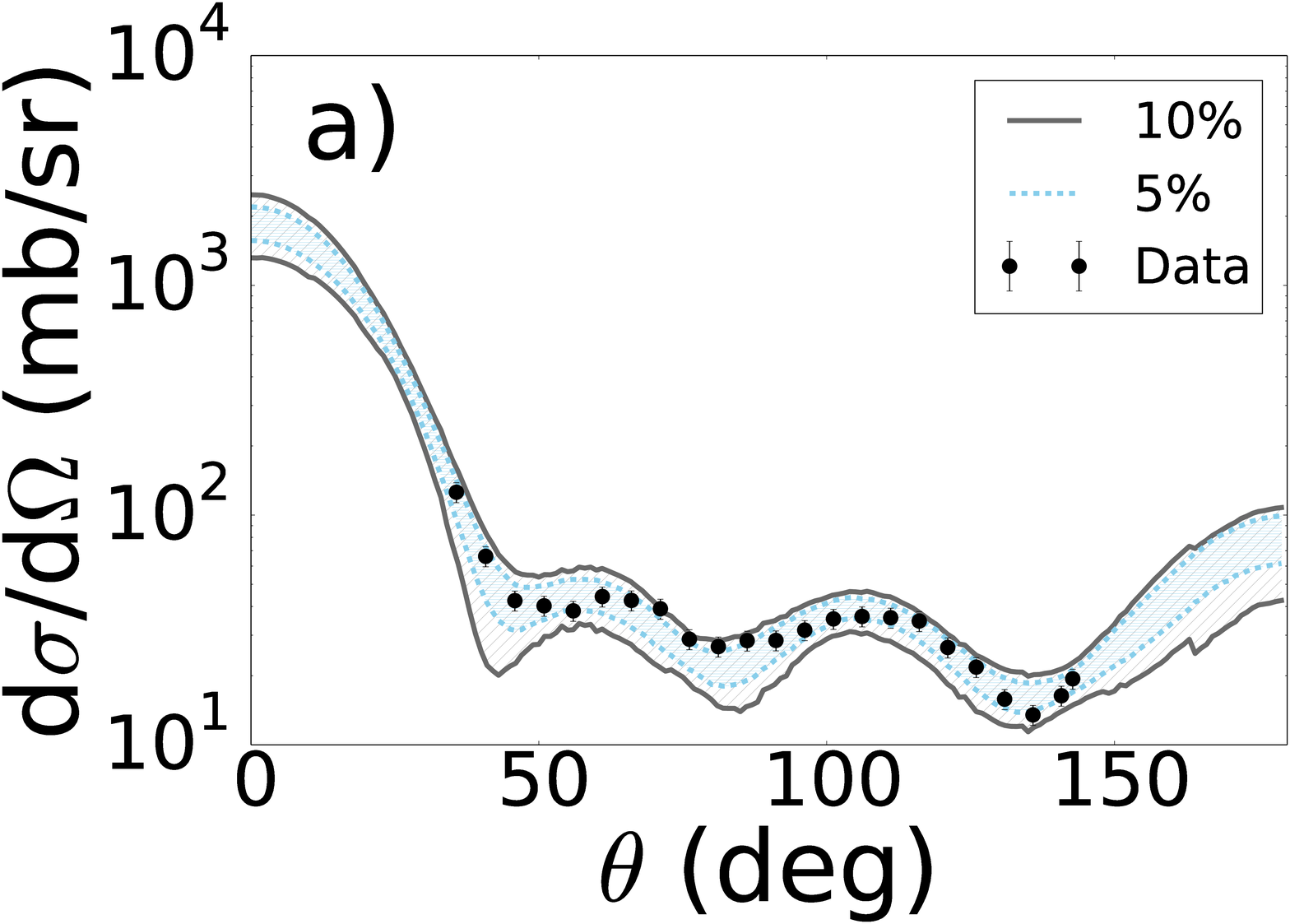} \\
\includegraphics[width=0.4\textwidth]{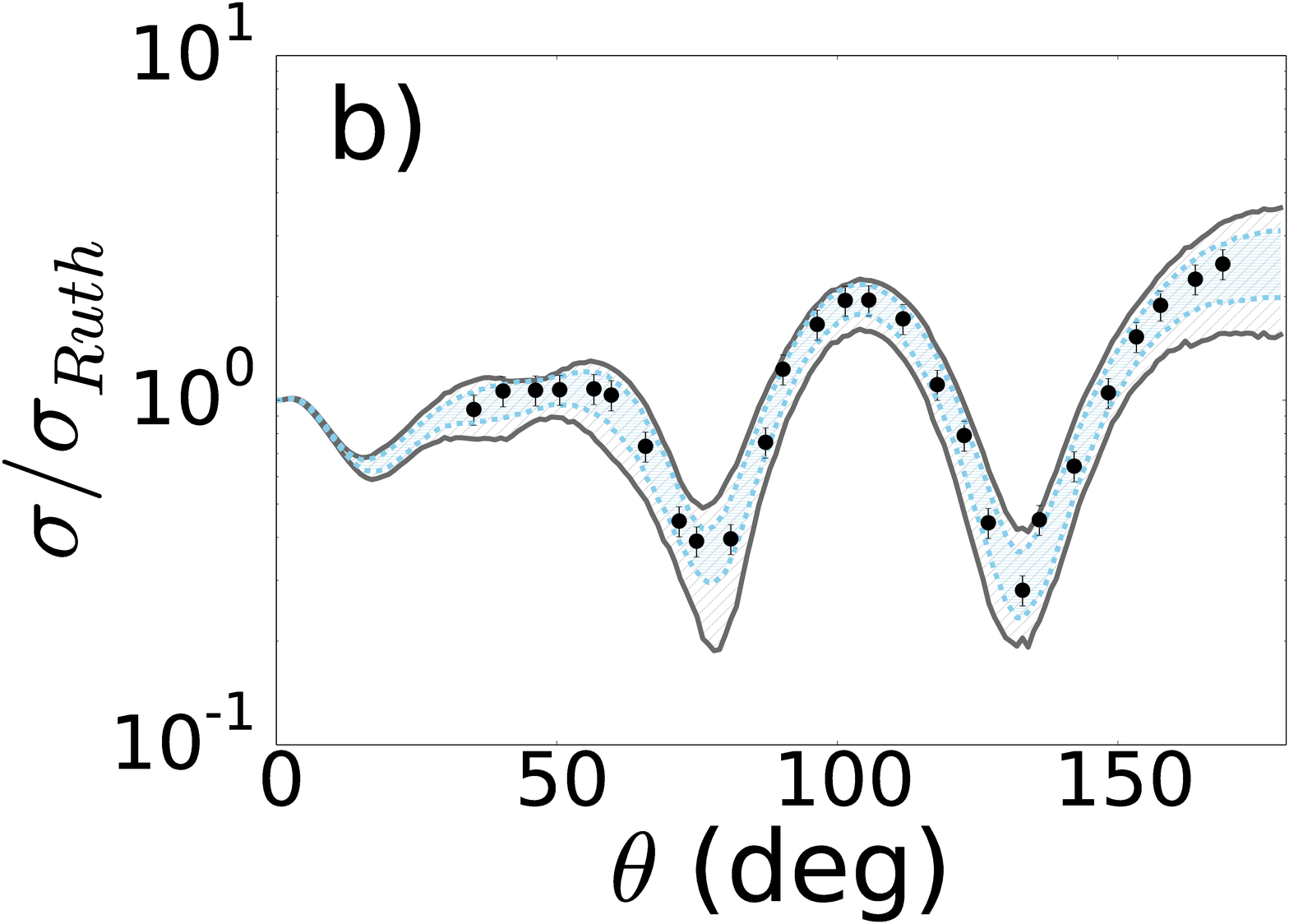} \\
\includegraphics[width=0.4\textwidth]{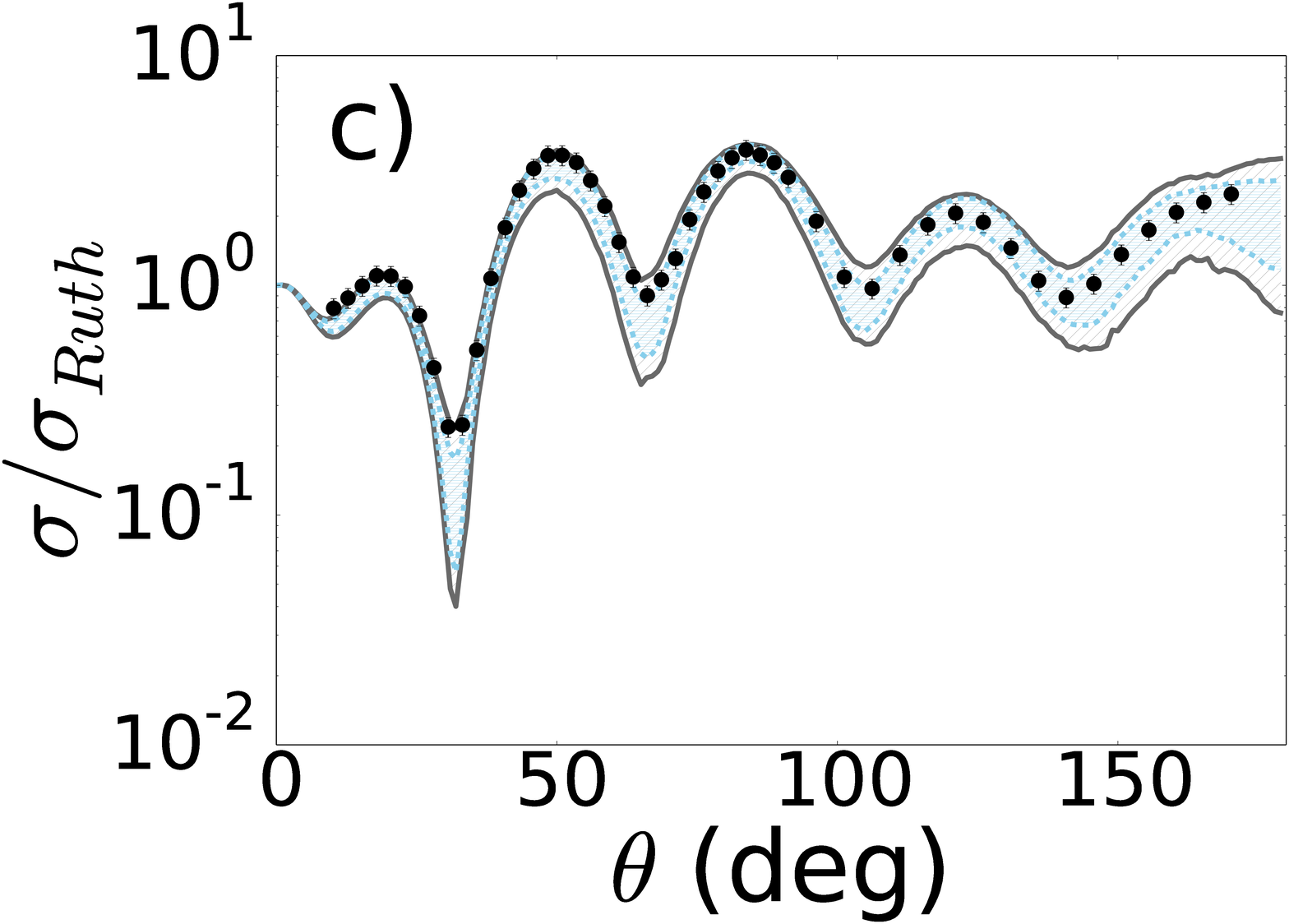} \\
\caption{(Color online) 95\% confidence intervals for the elastic scattering of a) $^{48}$Ca(n,n) at 12.0 MeV, b) $^{48}$Ca(p,p) at 14.08 MeV, and c) $^{48}$Ca(p,p) at 25.0 MeV.  Gray solid (light blue dashed) lines outline the 95\% intervals when 10\% (5\%) experimental errors were used.}
\label{fig:48CaADWAelastic}
\end{center}
\end{figure}

\begin{figure}
\begin{center}
\includegraphics[width=0.4\textwidth]{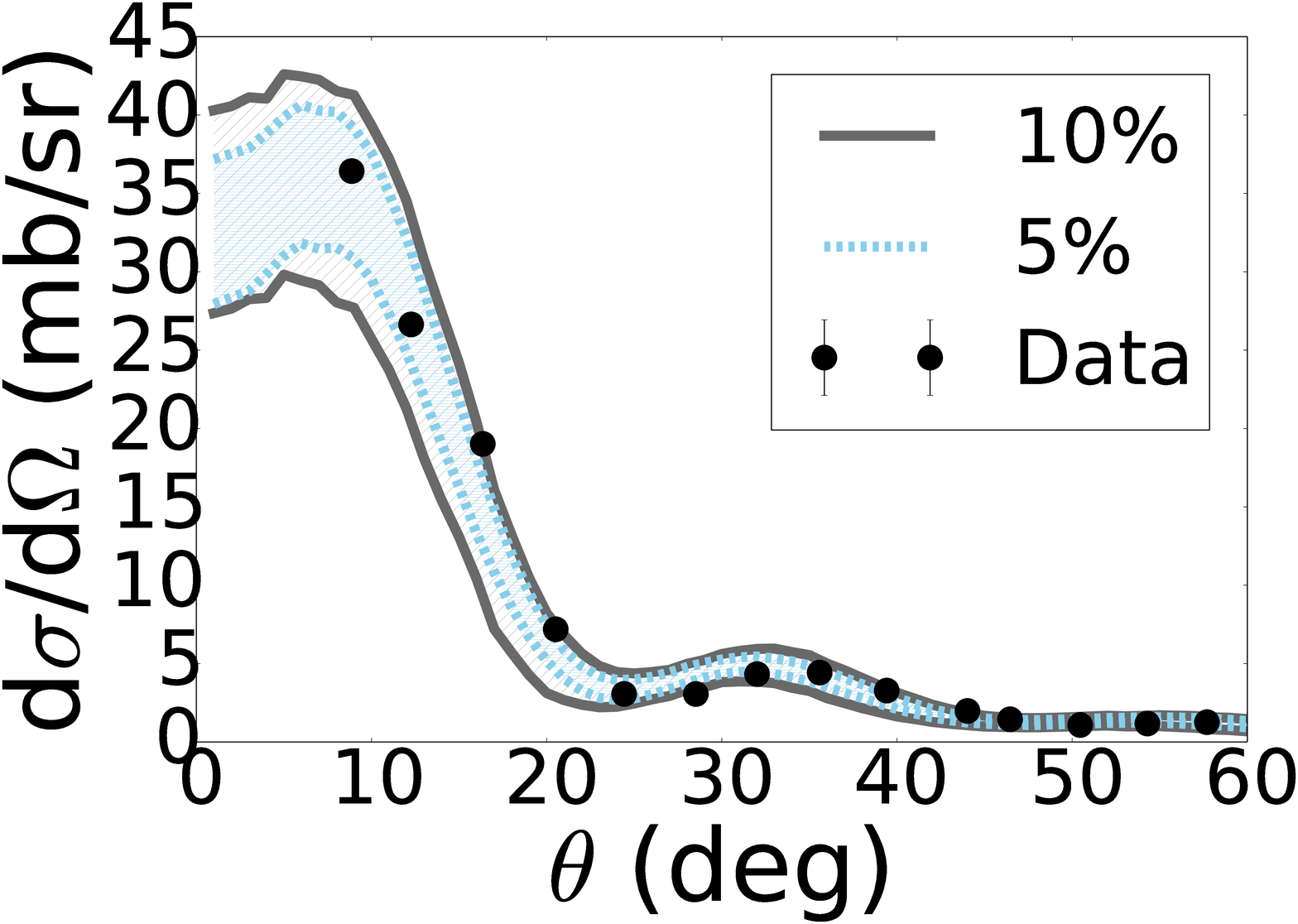}
\caption{(Color online) 95\% confidence interval for $^{48}$Ca(d,p)$^{49}$Ca(g.s.) at 24.0 MeV, compared to data at 19.3 MeV (extracted from \cite{48Cadp193}).  Gray solid (light blue dashed) regions show the intervals when 10\% (5\%) experimental errors are used for the ADWA calculation.}
\label{fig:48CaADWAdp}
\end{center}
\end{figure}

Figure \ref{fig:48CaADWAelastic} shows the 95\% confidence intervals (black) resulting from these parameter posterior distributions.  The data is well reproduced by the Monte Carlo sampling under the constraint of Bayes' Theorem, and the confidence intervals are relatively narrow.  The parameter sets that make up the posterior distributions for the elastic scattering channels can then be used to calculate the transfer cross section. For this purpose, we randomly draw three parameter sets, one from the incoming neutron posterior, one from the incoming proton posterior, and another from the outgoing proton posterior. These are then  combined to calculate the ADWA transfer cross sections.  It is important to remember that we use a fixed mean field to produce the single particle bound state in the exit channel. In order to mimic the introduction of a spectroscopic factor into our problem, we normalize the predicted angular distributions to the data at forward angles.  The 95\% confidence interval for this calculation, after normalization, is shown in Figure \ref{fig:48CaADWAdp} (black). 

We can then systematically study how the reduction of the experimental errors changes the resulting transfer calculation.  To do this, we rerun the Monte Carlo with the same prior distributions but assuming 5\% errors on the experimental data.  
The resulting mean parameter values are approximately the same but the widths are generally smaller - which can also be seen in Figures \ref{fig:48Can12}, \ref{fig:48Cap1403}, and \ref{fig:48Cap25} (blue histograms).  The $\chi^2$ values are larger when the experimental errors decrease; this is expected since the $\chi^2$ is weighted by the now smaller error at each angle.  Figure \ref{fig:48CaADWAelastic} shows the comparison of the 95\% confidence intervals for the elastic-scattering cross sections using these two errors (10\% errors in black and 5\% errors in blue).  As one would expect, the cross section confidence intervals are narrower when smaller error bars are used; to an extent, we can better constrain our calculations when the data is measured more precisely.  Finally, Figure \ref{fig:48CaADWAdp} compares the 95\% confidence intervals for the transfer reactions using the two posterior distributions. Reducing the error on the elastic cross section data by 50\%, reduces the uncertainty in the predicted transfer cross sections by $\approx 30$\%. We will come back to this issue in Sec. \ref{discussion}.

\subsubsection{Transfer using DWBA}
\label{DWBA}


We can perform the same study using the deuteron elastic scattering data to constrain the incoming channel (through DWBA) instead of the incoming nucleon interactions (with ADWA).  The prior for the deuteron-target elastic scattering also has the Gaussian form of Eq. (\ref{eqn:Gauprior}) for each parameter included in the fit, centered on the original optical potential values, now from \cite{ACOpt}, with a width equal to the center value.  (The outgoing proton or neutron channels are defined from the same posterior distributions as in the ADWA study.)  Similar posterior distributions are obtained (not shown), when using 10\% and 5\% error bars on the data, both in the mean values and widths; all of the parameter posteriors are centered around physical values.  Like the nucleon elastic scattering, for deuteron elastic scattering the imaginary volume diffuseness, $a_w$, could not be constrained by the data and therefore was not included in the fit.  It is fixed at its original value from \cite{ACOpt}.

\begin{figure}
\begin{center}
\includegraphics[width=0.4\textwidth]{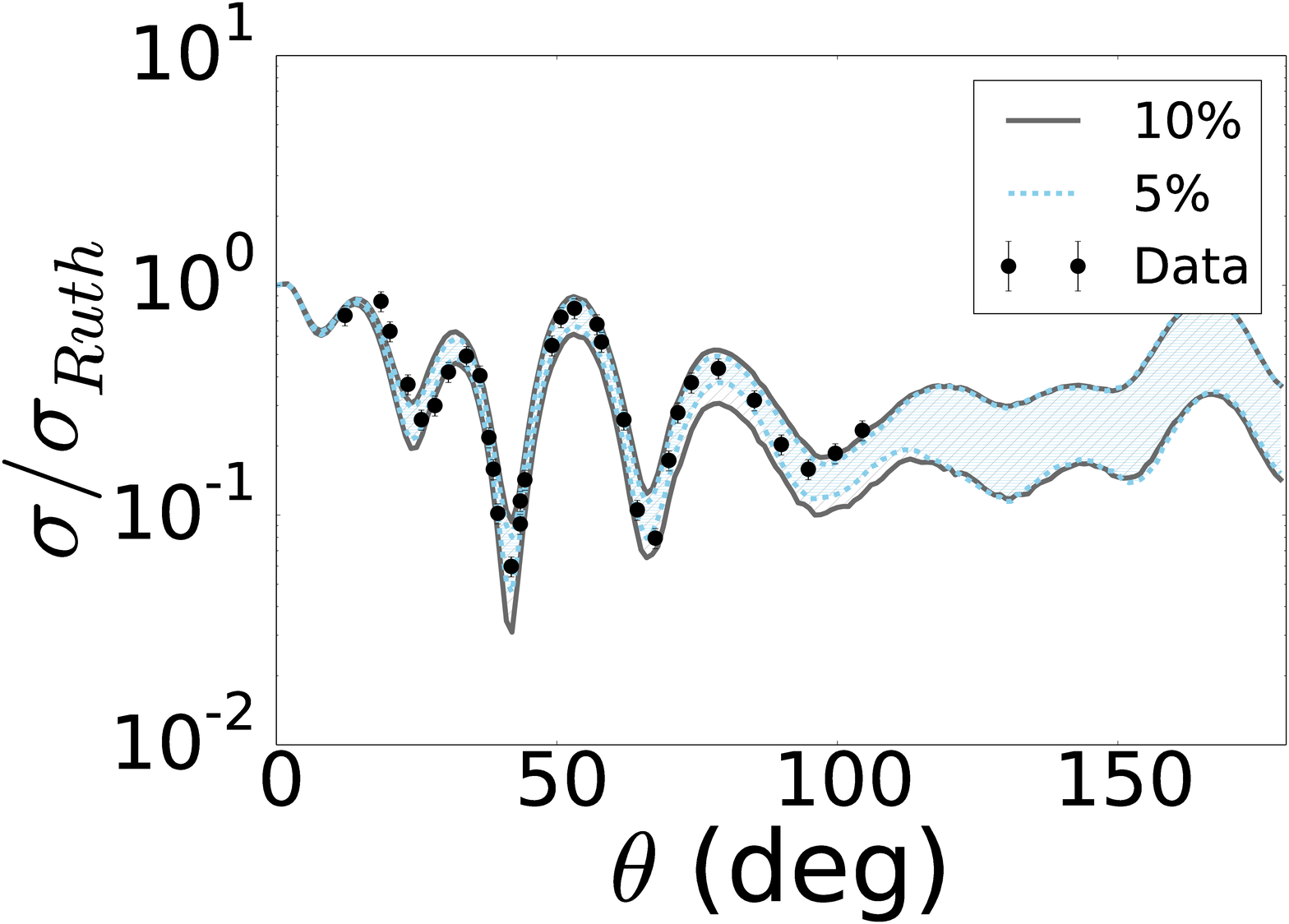}
\caption{(Color online) Same as Figure \ref{fig:48CaADWAelastic} for $^{48}$Ca(d,d)$^{48}$Ca at 23.2 MeV.}
\label{fig:48CaDWBAelastic}
\end{center}
\end{figure}

\begin{figure}
\begin{center}
\includegraphics[width=0.4\textwidth]{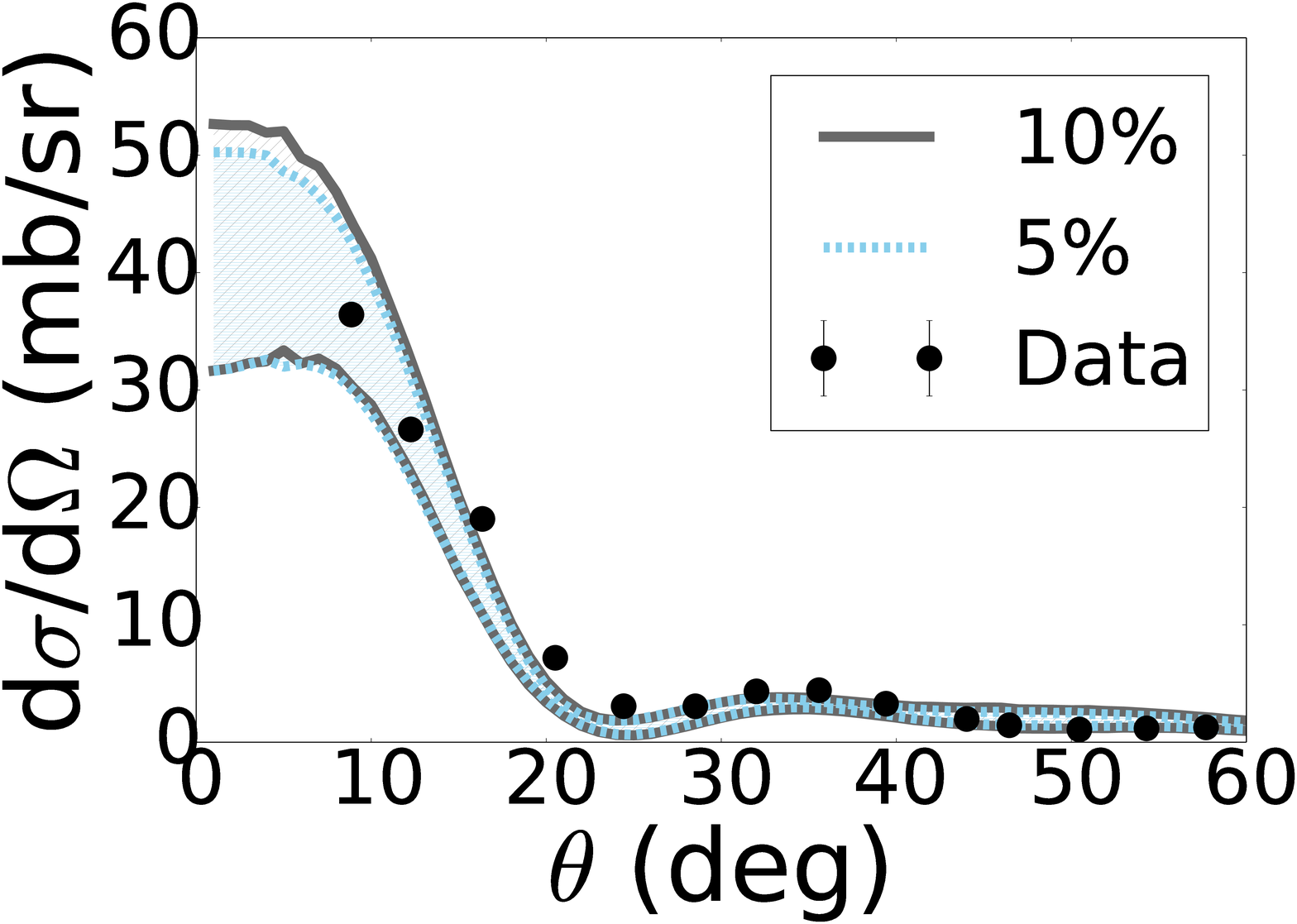} \\
\caption{(Color online) Same as Figure \ref{fig:48CaADWAdp} for the DWBA calculation.}
\label{fig:48CaDWBAdp}
\end{center}
\end{figure}

The 95\% confidence intervals for the deuteron elastic scattering are shown in Figure \ref{fig:48CaDWBAelastic}.  Again, we see that these intervals are well constrained based on the data, although the angular range covered by the data is significantly smaller than that covered by the nucleon scattering data.  We then use the deuteron elastic-scattering posterior and the outgoing proton elastic-scattering posterior shown in Figure \ref{fig:48Cap25} for the ADWA calculation (posteriors from Figure \ref{fig:48Cap25}) to perform the DWBA calculation.  Figure \ref{fig:48CaDWBAdp} shows the $^{48}$Ca(d,p)$^{49}$Ca(g.s.) transfer cross sections using the 10\% (black) and 5\% (blue) errors, using DWBA.  Here we see almost no reduction in the width of the transfer cross section at the peak when the smaller experimental errors are included.  The reduction is on the order of 10\% and only occurs at the peak.  

\subsubsection{Comparison between reaction models}

\begin{figure}
\begin{center}
\includegraphics[width=0.4\textwidth]{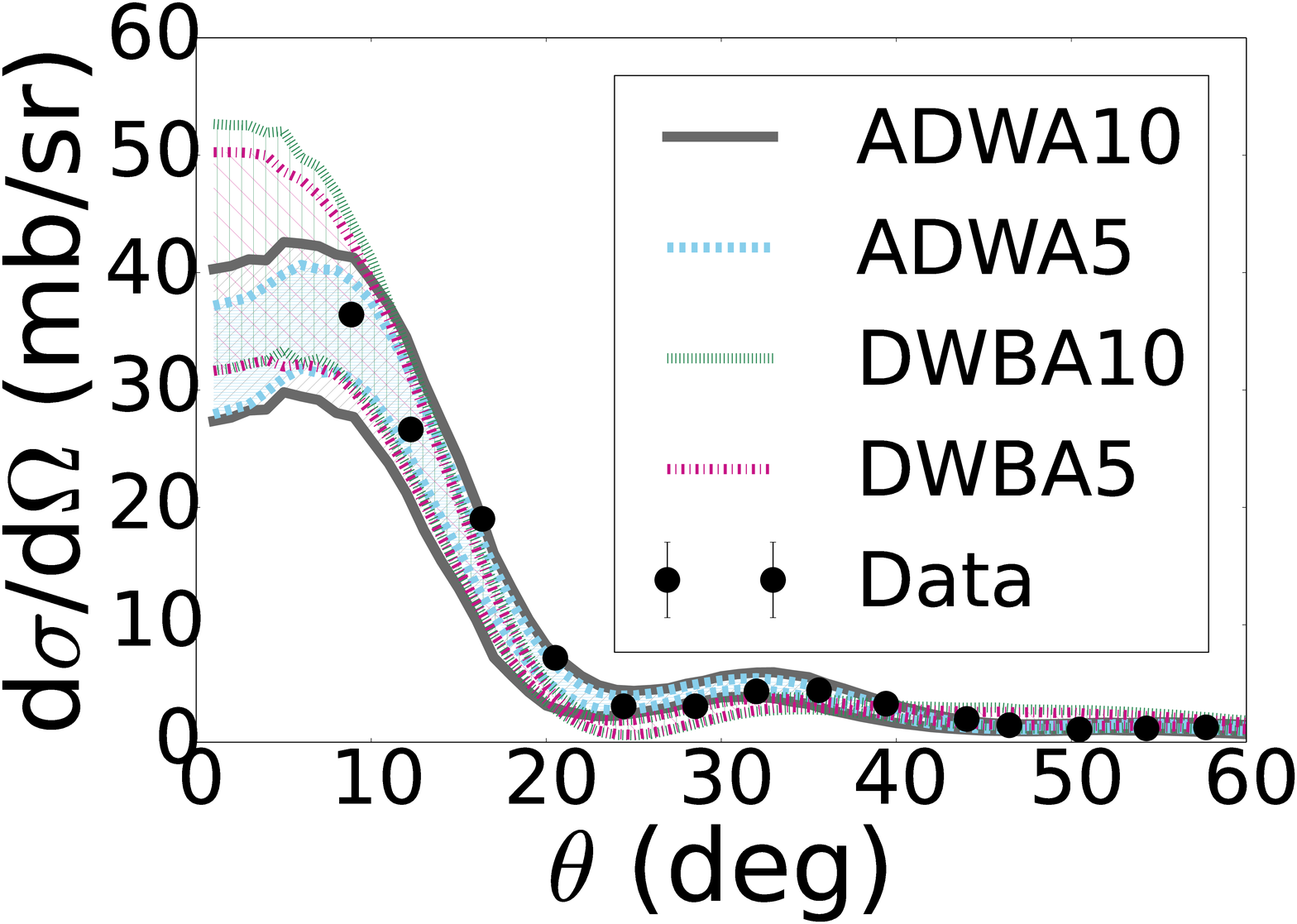} \\
\caption{(Color online) Comparison of angular distributions between ADWA with 10\% errors (gray solid), ADWA with 5\% errors (light blue dashed), DWBA with 10\% errors (green dash dotted), and DWBA with 5\% errors (pink dotted) for $^{48}$Ca(d,p)$^{49}$Ca(g.s.) at 24.0 MeV (compared to data at 19.3 MeV).}
\label{fig:percomp}
\end{center}
\end{figure}

We can now directly compare the two reaction models.  Figure \ref{fig:percomp} overlays the confidence bands obtained using ADWA  and DWBA, for  both 10\% and 5\% experimental errors, all normalized to the experimental data at forward angles.  The reaction models produce transfer cross sections that differ slightly in their angular dependence, although they all peak around the same angle - close to $5^\circ$ (this is mainly due to the same kinematic conditions and the same angular momentum transfer in ADWA and DWBA).  
At the peak of the angular distribution - where a spectroscopic factor would typically be extracted - the DWBA calculations have larger uncertainties than the ADWA calculations by $25-40\%$.  


\subsection{Summary of results}
\label{summary}



\begin{table}
\begin{center}
\begin{tabular}{ccccccc}
\hline \hline \textbf{Reaction} & \textbf{Theory} & \textbf{$\theta$} & \textbf{Peak$^*$} & \textbf{SF} & \textbf{$\varepsilon_{95}$} & \textbf{$\varepsilon_{68}$} \\ \hline
$^{48}$Ca(d,p) &  ADWA10 & 6 & 34.09 & 1.068 & 35.76 & 16.47 \\
$^{48}$Ca(d,p) &  ADWA5 & 6 & 33.38 & 1.092 & 24.24 & 11.53 \\
$^{48}$Ca(d,p) &  DWBA10 & 3 & 41.56 & 1.017 & 47.93 & 22.57 \\
$^{48}$Ca(d,p) &  DWBA5 & 4 & 40.73 & 1.016 &42.03 & 22.36 \\ \hline
$^{90}$Zr(d,n) &  ADWA10 & 31 & 2.16 & --- & 44.44 & 17.59 \\
$^{90}$Zr(d,n) &  ADWA5 & 31 & 2.13 & --- & 20.19 & 9.91 \\
$^{90}$Zr(d,n) &  DWBA10 & 31 & 3.04 & --- & 38.82 & 21.52 \\
$^{90}$Zr(d,n) &  DWBA5 & 30 & 3.15 & --- & 26.35 & 13.29 \\ \hline
$^{90}$Zr(d,p) &  ADWA10 & 14 & 16.63 & 0.740 & 47.62 & 21.95 \\
$^{90}$Zr(d,p) &  ADWA5 & 14 & 17.94 & 0.686 & 30.88 & 14.99 \\
$^{90}$Zr(d,p) &  DWBA10 & 16 & 17.09 & 0.720 & 58.86 & 29.02\\
$^{90}$Zr(d,p) &  DWBA5 & 16 & 17.41 & 0.707 & 30.61 & 14.26 \\ \hline
$^{116}$Sn(d,p) &  ADWA10 & 1 & 4.64 & --- & 121.77 & 48.31 \\
$^{116}$Sn(d,p) &  ADWA5 & 1 & 5.93 & --- & 101.52 & 55.12 \\ \hline
$^{208}$Pb(d,p) &  ADWA10 & 11 & 13.32 & --- & 37.84 & 18.95 \\
$^{208}$Pb(d,p) &  ADWA5 & 14 & 13.97 & --- & 25.48 & 11.42 \\
$^{208}$Pb(d,p) &  DWBA10 & 9 & 7.44 & --- & 72.72 & 43.84 \\
$^{208}$Pb(d,p) &  DWBA5 & 7 & 8.38 & --- & 63.01 & 30.08 \\ \hline \hline
\end{tabular}
\caption{Overview of the uncertainty in the differential cross section for  each transfer reaction.  Column one lists the transfer reaction, and column two lists the reaction theory used (ADWA or DWBA) with 5 or 10 indicating the experimental errors used.  The angle ($\theta$, in degrees) where the cross section peaks is listed in column three, and the value of the cross section at the peak (in mb/sr) is listed in column four.  The spectroscopic factors are given in column five (for the reactions that have been measured experimentally).  Column six (seven) lists the percentage error at the peak assuming a 95\% (68\%) confidence interval.}
\label{tab:summary}
\end{center}
\end{table}

For all but one transfer reaction calculated for this work, we follow the same procedure as in Section IV C and compile the results in this section (the DWBA calculation was not performed for $^{116}$Sn(d,p)$^{117}$Sn due to a lack of $(d,d)$ scattering data at the incident energy).  
In Table \ref{tab:summary}, we present the widths of the confidence bands  predicted for transfer reactions calculated in this work.  It lists the mean values, at the peak of the angular distributions for the 95\% confidence intervals (column four), given a reaction model (ADWA or DWBA), with the index representing the experimental error taken for the elastic scattering cross sections (5 and 10 for 5\% errors and 10\% errors, respectively).  Note that the peak values are those corresponding to the 95\% confidence intervals.  These values may change by 5-10\% at most when 68\% confidence intervals are calculated, but fall within the 95\% intervals.  Two percentage uncertainty widths are listed in columns five and six are defined as
\begin{equation}
\label{eqn:thUQ}
\varepsilon_i = \frac{\sigma ^i_\mathrm{max} - \sigma^i_\mathrm{min}}{\bar{\sigma}^i} \times 100\%\;,
\end{equation}

\noindent where $i$ indicates  which confidence interval is being calculated (95\% or 68\%, as given by Eq. \ref{eqn:CIdef}), $\sigma^i_\mathrm{max}$ ($\sigma^i_\mathrm{min}$) give the maximum (minimum) values of the cross sections defined by the $i\%$ confidence interval, and $\bar{\sigma}^i$ denotes the mean value of the cross section at the peak within the $i\%$ confidence interval.  In principle, the 95\% confidence intervals should be about twice as wide at the 68\% confidence intervals.  When this does not hold, we can make inferences about the tails of the confidence intervals - whether or not they are asymmetric around the mean or how far they extend from the mean (equivalently, how peaked the distribution is).  In Table \ref{tab:summary}, we see that $\varepsilon _{68} \approx 0.5 \varepsilon _{95}$ for nearly every reaction indicating that the distributions are nearly symmetric and could be well described as Gaussian.

\begin{table}
\begin{center}
\begin{tabular}{cccc}
\hline \hline \textbf{Reaction} & \textbf{Theory} & \textbf{$(\Delta_{exp})_{95}$} & \textbf{$(\Delta_{exp})_{68}$} \\ \hline
$^{48}$Ca(d,p) &  ADWA & 32.22 & 30.03 \\
$^{48}$Ca(d,p) &  DWBA & 12.30 & 0.91 \\ \hline
$^{90}$Zr(d,n) &  ADWA & 54.58 & 43.69 \\
$^{90}$Zr(d,n) &  DWBA & 32.12 & 38.24 \\ \hline
$^{90}$Zr(d,p) &  ADWA & 35.15 & 31.68 \\
$^{90}$Zr(d,p) &  DWBA & 47.99 & 50.87 \\ \hline
$^{116}$Sn(d,p) &  ADWA & 16.63 & -14.10 \\ \hline
$^{208}$Pb(d,p) &  ADWA & 32.65 & 39.71 \\
$^{208}$Pb(d,p) &  DWBA & 13.35 & 31.39 \\ \hline \hline
\end{tabular}
\caption{Overview of the reduction (or increase) factor between the 10\% error calculations and the 5\% error calculation for the reaction model listed in column two.  This is done for both the 95\% and 68\% confidence intervals.  See text for details.}
\label{tab:differrors}
\end{center}
\end{table}

\begin{table}
\begin{center}
\begin{tabular}{cccc}
\hline \hline \textbf{Reaction} & \textbf{Error} & \textbf{$(\Delta_{th})_{95}$} & \textbf{$(\Delta_{th})_{68}$} \\ \hline
$^{48}$Ca(d,p) &  10\% & 25.39 & 27.03 \\
$^{48}$Ca(d,p) &  5\% & 42.33 & 48.43 \\ \hline
$^{90}$Zr(d,n) &  10\% & -14.48 & 18.26 \\
$^{90}$Zr(d,n) &  5\% & 23.38 & 25.43 \\ \hline
$^{90}$Zr(d,p) &  10\% & 19.10 & 24.36 \\
$^{90}$Zr(d,p) &  5\% & -0.88 & -5.12 \\ \hline
$^{208}$Pb(d,p) &  10\% & 47.96 & 56.77 \\
$^{208}$Pb(d,p) &  5\% & 147.29 & 62.03 \\ \hline \hline
\end{tabular}
\caption{Overview of the reduction (or increase) factor between the DWBA calculations and the ADWA calculation for the percent error on the experimental data listed in column two.  This is done for both the 95\% and 68\% confidence intervals.  See text for details.}
\label{tab:difftheory}
\end{center}
\end{table}


In order to assess the gain of predictive power when increasing the precision of the experimental data, we introduce:
\begin{equation}
\Delta_{exp} = \frac{\varepsilon_i(\mathrm{ADWA10}) - \varepsilon_i(\mathrm{ADWA5})}{\varepsilon_i (\mathrm{ADWA10})} \times 100\%.
\end{equation}
In Table \ref{tab:differrors}, we show this $\Delta_{exp}$ factor, for each theory model.  

Also of interest is the information gain when improving the theoretical description of the reaction. As mentioned earlier, ADWA is built on a three-body model for the reaction and contains deuteron breakup to all orders, while the standard DWBA calculations consist of the first term of a perturbative series based on two-body multiple scattering. We thus introduce the quantity:
\begin{equation}
\Delta_{th} = \frac{\varepsilon_i(\mathrm{DWBA}) - \varepsilon_i(\mathrm{ADWA})}{\varepsilon_i (\mathrm{DWBA})} \times 100\%
\end{equation}
which reflects the improvement in describing the data when taking deuteron breakup explicitly.
This is shown in Table \ref{tab:difftheory}. Because the DWBA calculation was not performed for $^{116}$Sn, $\Delta_{th}$ could not be calculated for this system.

Overall we find that reducing experimental errors and improving the reaction theory, reduces the uncertainty in the prediction of the transfer cross section. In the next section, we present a thorough discussion; however we should underline two atypical results.
For $^{116}$Sn(d,p)$^{117}$Sn, we find that the width of the 68\% confidence interval  increases when the experimental error is halved (see Table \ref{tab:differrors}).  This is due to the non-symmetric nature of the posterior distribution of the transfer cross section.  For the 68\% confidence intervals, there is no strong peak in the density of the cross section values at each angle, which results in the mean being defined rather arbitrarily. The other unusual case is $^{90}$Zr(d,n)$^{91}$Nb for which there is an increase  in the theoretical uncertainty for the ADWA calculations compared to the DWBA calculations (see Table \ref{tab:difftheory}). It turns out that the deuteron optical potential of \cite{ACOpt} seems to be particularly well-defined for $^{90}$Zr-d scattering, especially compared to $^{48}$Ca and $^{208}$Pb.  In \cite{ACOpt}, the authors quote an overall $\chi^2$ of 4.03 for $^{90}$Zr over the range of energies that were investigated ($E<183$ MeV) which contrasts with $^{48}$Ca, $\chi^2 = 123.58$. This is the reason why in some particular combination of confidence level and experimental error bar precision, DWBA appears to perform slightly better. 

\section{Discussion}
\label{discussion}


The three main goals of our analysis are:  1) assess the uncertainties in the transfer cross section, within the adiabatic formalism, when constraining the nucleon-target optical potentials with the relevant elastic scattering data, 2) understand the effect of data precision on the resulting cross section uncertainties, and 3) systematically compare the two reaction formalisms which introduce different approximations.  These are each discussed in this section.

\subsection{Uncertainties from nucleon-target potentials}
\label{sec:NApot}

We can now examine the uncertainties from the nucleon-target interactions.  If we first focus on the uncertainties from the 95\% confidence intervals (column six of Table \ref{tab:summary}) for the ADWA calculations, we see that these range from 20\% to about 120\%.  Almost all of these are larger than the 10\% to 30\% that is na\"ively expected to come from the parameterization of the optical model.  (We see that the uncertainties of the 95\% confidence intervals for the DWBA calculations are larger on average; these differences will be discussed more in Section \ref{sec:ADWADWBA}.)

\begin{table}
\begin{center}
\begin{tabular}{ccccc}
\hline \hline \textbf{Reaction} & \textbf{Projectile} & \textbf{$\varepsilon_{10}$} & \textbf{$\varepsilon_5$} & \textbf{$\theta_{max}$ (deg)} \\ \hline
$^{48}$Ca(d,p) & p$_{in}$ & 22.90 & 11.82 & 158 \\ 
$^{48}$Ca(d,p) & n$_{in}$ & 15.82 & 7.96 & 143 \\
$^{48}$Ca(d,p) & p$_{out}$ & 26.70 & 17.37 & 170 \\
$^{48}$Ca(d,p) & d$_{in}$ & 26.08 & 15.61 & --- \\ 
$^{48}$Ca(d,p) & AD$_{\mathrm{quad}}$ & 38.57 & 22.47 & --- \\ \hline
$^{90}$Zr(d,n) & p$_{in}$ & 18.44 & 15.28 & 165 \\
$^{90}$Zr(d,n) & n$_{in}$ & 16.96 & 9.17 & 150 \\
$^{90}$Zr(d,n) & n$_{out}$ & 26.04 & 12.08 & 159 \\
$^{90}$Zr(d,n) & d$_{in}$ & 28.72 & 17.17 & --- \\
$^{90}$Zr(d,n) & AD$_{\mathrm{quad}}$ & 36.14 & 21.53 & --- \\ \hline
$^{90}$Zr(d,p) & p$_{in}$ & 17.53 & 12.81 & 165 \\
$^{90}$Zr(d,p) & n$_{in}$ & 13.78 & 8.92 & 150\\
$^{90}$Zr(d,p) & p$_{out}$ & 38.24 & 19.96 & 154 \\
$^{90}$Zr(d,p) & d$_{in}$ & 23.77 & 19.18 & --- \\
$^{90}$Zr(d,p) & AD$_{\mathrm{quad}}$ & 44.27 & 25.34 & --- \\ \hline
$^{116}$Sn(d,p) & p$_{in}$ & 80.50 & 64.60 & 169 \\
$^{116}$Sn(d,p) & n$_{in}$ & 35.26 & 18.43 & 155 \\
$^{116}$Sn(d,p) & p$_{out}$ & 87.05 & 79.64 & 88 \\
$^{116}$Sn(d,p) & d$_{in}$ & 88.65 & 64.16 & --- \\
$^{116}$Sn(d,p) & AD$_{\mathrm{quad}}$ & 123.70 & 104.19 & --- \\ \hline
$^{208}$Pb(d,p) & p$_{in}$ & 16.42 & 7.76 & 165 \\
$^{208}$Pb(d,p) & n$_{in}$ & 22.35 & 12.92 & 154 \\
$^{208}$Pb(d,p) & p$_{out}$ & 33.00 & 21.41 & 168 \\
$^{208}$Pb(d,p) & d$_{in}$ & 30.98 & 19.62 & --- \\ 
$^{208}$Pb(d,p) & AD$_{\mathrm{quad}}$ & 43.11 & 26.69 & --- \\ \hline \hline
\end{tabular}
\caption{Theoretical uncertainties, $\varepsilon_i$, using 10\% (5\%) experimental errors, extracted at the peak of the cross section, column three (four) for a given transfer reaction (column one).  The projectile in column two indicates which part of the potential was varied (while the remaining nucleon-target potentials were fixed at the original parameterizations from \cite{BGOpt}.  (Here, the error on the deuteron channel comes from varying the incoming neutron and proton potentials simultaneously, and AD$_{\mathrm{quad}}$ comes from adding the errors from the nucleon incoming and outgoing potentials in quadrature.)  Column five shows the largest angle at which the experimental data was measured.}
\label{tab:parts}
\end{center}
\end{table}

Table \ref{tab:parts} shows the theoretical uncertainty as defined in Eq. \ref{eqn:thUQ} - at the peak of the transfer cross section - that is introduced when only one or two of the nucleon-target potential posterior distributions are included in the ADWA transfer calculation.  For nearly all of the calculations, the largest single channel uncertainty is introduced by the outgoing nucleon-target potential.  This is an intriguing results that is not yet fully understood.  All of the scattering pairs (besides $^{116}$Sn(p,p) in the outgoing channel) have data out to $150^\circ$ or beyond, so we cannot attribute this result to the lack of angular coverage.\footnote{For $^{116}$Sn case, the incoming and outgoing proton scattering for $^{116}$Sn have nearly the same percentage error, but the data for the proton in the incoming channel covers a significantly larger range.}  This result suggests a significant change in the way deuteron induced transfer reactions are currently measured. Typically the reaction is measured in inverse kinematics with a deuterated target.  We propose that in addition, a proton target is used to capture the elastic with the beam energy adjusted to match the relevant outgoing channel kinematic conditions. This will minimize the uncertainty coming from the optical potentials in the theoretical prediction. 

Table \ref{tab:parts} also lists the total quadrature uncertainty from the incoming and outgoing channels for the ADWA calculations, $\mathrm{AD}_{\mathrm{quad}}$.  We find that this quadrature uncertainty is nearly identical to the uncertainties in Table \ref{tab:summary}.  The total uncertainty calculated by including the uncertainties from all potentials simultaneously is the same as adding in quadrature the uncertainties from each potential individually.  We find the same results for the quadrature uncertainties for DWBA.

\subsection{Effects of the experimental error}

In Figures \ref{fig:48CaADWAdp} and \ref{fig:48CaDWBAdp}, we showed the comparison between the uncertainty for the ADWA and DWBA calculations using 10\% and 5\% errors on the experimental data for the case of $^{48}$Ca(d,p)$^{49}$Ca(g.s.).  In both cases, we see a reduction in the width of the cross section at the first peak (where a spectroscopic factor would be extracted), when smaller errors are included.  However, this reduction is not proportional to the reduction in the experimental error - for a 50\% reduction of the experimental error, we only find about a 30\% reduction in the uncertainty of the resulting transfer cross section for the ADWA calculation and a reduction of about 10\% for the DWBA calculation.  

Since we have used an exponentiated $\chi^2$ as our likelihood, one might na\"ively argue that a reduction in the experimental error by two should just scale the likelihood and therefore just scale the posterior distributions.  However, this is not what is obtained for our calculations.  We also do not see a consistent scaling of the widths of the parameter posterior distributions.  

As we saw for the elastic nucleon-target and deuteron-target cross sections used in the $^{48}$Ca(d,p)$^{49}$Ca(g.s.), in all other cases here studied, reducing the experimental error bars gives rise to similar means but smaller widths for the posterior distributions. This also has the effect of reducing the width of the 95\% confidence intervals for the elastic scattering (similar to what is seen in Figures \ref{fig:48CaADWAelastic} and \ref{fig:48CaDWBAelastic}).  The reduction in the width is not always drastic, but it is always present.  Further, we always see a reduction in the percentage error at the peak of the cross section when the experimental errors are halved (Table \ref{tab:differrors}), except for the 68\% confidence intervals for $^{116}$Sn as discussed in Section \ref{summary}.  This reduction ranges from about 10\% to 55\%.  Therefore, while tighter constraints on the experimental data allow us to more precisely extract information from the transfer cross section, the magnitude of the improvement in the prediction is not trivially related to the magnitude of improvement in the experimental measurement.  

\subsection{Comparison of ADWA and DWBA}
\label{sec:ADWADWBA}


In Table \ref{tab:difftheory}, we showed the reduction in the uncertainties for the transfer cross sections when going from a DWBA description to  ADWA. This is done for the two experimental error bars considered and for both the 95\% and 68\% confidence intervals.  Except for those cases discussed earlier, there is a significant reduction in the percentage error, at the peak, when improving the physics of the model. On average the width of the 95\% confidence band is $\approx 40\%$ for ADWA, and $\approx 55$\% for DWBA when 10\% error bars are included in the experiment. If instead 5\% error bars are considered, then the average uncertainty for the 95\% confidence band is $\approx 25$\% for ADWA and $\approx 40$\% for DWBA. This result confirms expectations, especially because the ADWA method has been shown to reproduce the exact solution of the three-body problem for the energies of interest in this work \cite{Nunes2011}.  As shown in Figure \ref{fig:percomp}, our ADWA angular distributions appear more in line with the experimental angular distribution but DWBA cannot be ruled out  due to the model uncertainties.  Although the ADWA and DWBA calculations have different peak shapes (data at more forward angles could distinguish between the calculations), the calculations at backwards angles do not provide the same differentiation.


The fifth column of Table \ref{tab:summary} lists the spectroscopic factors for $^{48}$Ca(d,p) and $^{90}$Zr(d,p) (which were the only two reactions where (d,p) data was available, from \cite{48Cadp193} and \cite{90Zrdp22}, respectively).  The spectroscopic factors were calculated by normalizing the mean theoretical cross section at the peak of the experimental angular distribution or the forward-most measured data point.  For each reaction, the spectroscopic factors between ADWA and DWBA differ by only a few percent which is significantly smaller than the uncertainty introduced by the optical potentials.  The differences in the spectroscopic factors alone would not be enough to distinguish between the two models especially considering the relatively large uncertainties that are introduced by the free parameters within the potential model.  

One should keep in mind that, as discussed earlier in Section \ref{sec:NApot}, these uncertainties are only due to the optical potential parameterizations, and the results may depend on the specifics of each set of data.  Ideally, we would like to have the same angular coverage for all relevant elastic scattering and transfer data.  Since ADWA and DWBA often predict different transfer angular distributions, such a study could enable model exclusion.  


\section{Conclusions and future work}
\label{conclusions}


We have used Bayesian methods to construct 95\% confidence intervals for five transfer reactions and relevant elastic scattering in the range $A=48-208$ with energies from 10 to 25 MeV/u.  The aim of the study is to quantify the uncertainties coming from the parameterization of the optical model potentials and to begin to investigate the uncertainties coming from differences in the reaction model implemented.  Nucleon-target and deuteron-target elastic scattering data were used to constrain the parameters in the potential and create posterior distributions for each parameter.  These posterior distributions were used to predict proton and neutron transfer cross sections, taking for the reaction formalism either the adiabatic wave approximation or the distorted-wave Born approximation.  The experimental errors for each data set were defined systematically to be a fixed percentage for all angles and data sets: we consider both  10\% and 5\% experimental errors. This enables a rigorous study on their impact in the  confidence intervals of the theoretically predicted observables.  

Overall, we find about $20-120\%$ error being introduced to the transfer angular distributions using data to constrain optical model parameters.  The uncertainties from each two-body scattering reaction essentially add in quadrature to produced the overall uncertainty when the potentials of all scattering pairs are varied simultaneously within the constraints of their posterior distributions.  The outgoing nucleon-$(A+1)$ potential introduces the largest uncertainty in the five cases studied here.  Reducing the experimental errors in the data significantly reduces the uncertainty in the constrained elastic and predicted transfer cross sections, however this effect is not directly proportional to the reduction factor of experimental error bars.  Finally, constraining the nucleon-target interactions and calculating a transfer cross section using ADWA generally introduces less uncertainty than constraining the deuteron-target interaction and predicting the transfer through DWBA.  We expect that ADWA would have less uncertainty as it explicitly takes the breakup of the deuteron into account.

Even though we have constrained each of the incoming and outgoing potentials with data and have included these in the overall uncertainties, there are still other uncertain elements in the theory, including the mean field potential binding the nucleon-target system in the final state and the $np$ interaction binding the deuteron in the initial state.  These uncertainties should also be quantified in the future.  

This work focused on the uncertainties due to the parameterization of the potentials. Given that the reactions here considered are many-body complex scattering problems, for which we use few-body methods, there is  the larger issue of model simplifications.  The ultimate goal is to estimate the uncertainties that arise from the model simplifications without knowing the exact solution, as well as rigorously comparing models to understand the information content and to what extent the increased complexity is justified by the evidence.
The Bayesian reaction framework we have implemented provides a way forward in this investigation.


\begin{center}
\textbf{ACKNOWLEDGMENTS}
\end{center}

The authors would like to thank David Higdon, Daniel Phillips, and Dick Furnstahl for their insightful discussions.  They would also like to acknowledge iCER and the High Performance Computing Center at Michigan State University for computational resources.  This work was supported by the Stewardship Science Graduate Fellowship program under Grant No. DE-NA0002135, the National Science Foundation under Grant. No. PHY-1403906,
and the Department of Energy under Contract No. DE-FG52-08NA28552.  


\appendix

\section{Linear Priors}
\label{app:priors}

\begin{figure}[h!]
\begin{center}
\epsfig{file=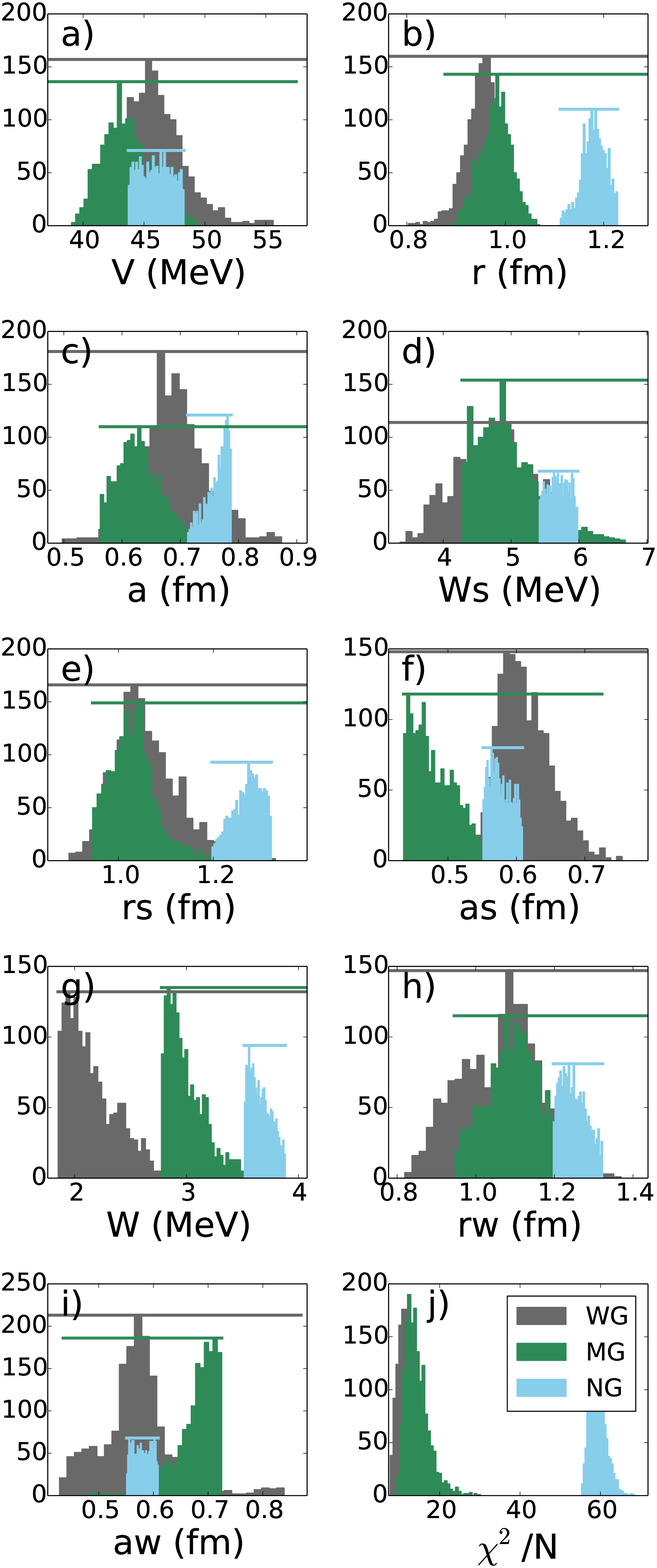,width=0.4\textwidth} 
\caption{(Color online)  Comparison of the posterior distributions (histograms) resulting from various prior distributions (corresponding solid lines) for a wide linear (WL), medium linear (ML), and narrow linear (NL) as defined in Table \ref{tab:priorshape} for $^{90}$Zr(n,n)$^{90}$Zr at 24.0 MeV.}
\label{fig:priorlinear}
\end{center}
\end{figure}

\begin{figure}[h]
\begin{center}
\epsfig{file=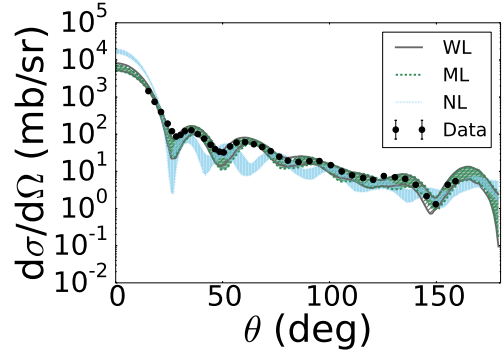,width=0.4\textwidth}
\caption{(Color online)  95\% confidence intervals of the elastic-scattering angular distributions for $^{90}$Zr(n,n)$^{90}$Zr at 24.0 MeV using the posterior distributions from Figure \ref{fig:priorlinear}.}
\label{fig:xslinear}
\end{center}
\end{figure}

Although we discussed the linear priors in the main text, they were not shown in detail.  Therefore, in Figure \ref{fig:priorlinear}, we show a comparison of the posterior distributions resulting from the linear priors listed in Table \ref{tab:priorshape}, considering again $^{90}$Zr(n,n)$^{90}$Zr at 24.0 MeV.  Contrary to the Gaussian priors of Section \ref{sec:priors}, in the case of the linear priors, the posterior distributions  sharply terminate at the boundaries of the prior for nearly all of the variables, particularly for the narrow linear (NL) prior.  Although this is not seen in the medium and wide prior for the real volume parameters, there is a significant effect on the posterior distributions for the imaginary terms.  This has a large impact on the resulting 95\% confidence intervals for the elastic-scattering angular distributions, as shown in Figure \ref{fig:xslinear}.  Despite the stark differences in the posterior distributions for the medium and wide  linear priors, the resulting angular distributions are strikingly similar.  The same is not true for the angular distribution resulting from the narrow prior which does not even reproduce the experimental data.

\bibliography{BayesBib}

\end{document}